%  This file has the file  of the paper: Grand Canonical Partition 
% Function of the Unidimensional Hubbard Model 
% up to Order beta^3, by i.c. charret, s.m. de souza, m.t. thomaz
% and  e.v. correa silva.The file is in Plain Tex.
% Please acknowledge receiving this file: mtt@if.uff.br
% Maria Teresa Thomaz (for the authors)

\magnification=1200
\pageno=0
\hsize=5.6in
\vsize=7.5in
\voffset=0.2in 
\hoffset= -0.2in

%\font\titulo1=cmb10 scaled 1200
\font\titulo=cmb10 scaled 1440 

%\phantom{...} 
%\bigskip 

% definicao do circulo com a letra interna, para representar as
% funcoes normais de Grassmann
\def\normal{{\bigcirc \!\hskip -5pt n}} 

%definicao da matrix 0
\def\0barra{{\rm O} \!\hskip -3.7pt {\rm l} } 

%definicao da matriz identidade
\def\lbarra{1\! \hskip -1.1pt {\rm l}} 

%definicao da letra de 7pontos em italico para colocar indices

\baselineskip=18pt
\centerline
{\titulo Grand Canonical Partition Function for Unidimensional
Systems:} \par 

\centerline{\titulo Application to Hubbard Model up to Order $\beta^3$}
			 \par 

\baselineskip=12pt
\vskip 1cm
\centerline {I.C. Charret\footnote{$^1$}{E--mail: iraziet@if.uff.br}, 
S.M.  de Souza\footnote{$^2$}{E--mail: smartins@if.uff.br},
 M.T. Thomaz\footnote{$^3$}{Corresponding author: Dr. Maria Teresa Thomaz;
 R. Domingos S\'avio Nogueira Saad n.$\!\!^{\rm o}$ 120  apto 404, 
 Niter\'oi, R.J., 24210--340, BRAZIL
 --Phone/Fax: (21) 620--6735; \hfill \break E--mail: mtt@if.uff.br }
} \par
\vskip 0.3cm
\centerline {\it Instituto de F\'\i sica} \par
\centerline {\it Universidade Federal Fluminense} \par
\centerline {\it R. Gal. Milton Tavares de Souza s/n.$\!\!^o$} \par
\centerline{\it Campus da Praia Vermelha} \par
\centerline {\it Niter\'oi, R.J., 24210--310 $\quad$ BRAZIL} \par 

\vskip 0.6cm 

\centerline{ E.V. Corr\^ea Silva\footnote{$^4$}{E--mail:
ecorrea@cbpfsu1.cat.cbpf.br }} \par
\vskip 0.3cm
\centerline{\it Centro Brasileiro de Pesquisas F\'{\i}sicas} \par
\centerline{\it R. Dr. Xavier Sigaud n.$\!\!^{\rm o}$ 150} \par
\centerline{\it Rio de Janeiro, R.J., 22290-180 $\quad$ BRAZIL } \par

\baselineskip=14pt

\vskip 1cm

\centerline {\bf Abstract} \par

\bigskip 

We exploit the grassmannian nature of the variables involved in the
path integral expression of the grand canonical partition function for
self--interacting fermionic models to show, in one--space dimension,
a general relation among the terms of it expansion in the high 
temperature limit and a combination of co--factors of a
suitable matrix with commuting entries. As an application, we apply 
this framework to calculate the exact coefficients, up to
 order $\beta^3$, of the expansion of the  grand canonical partition
function for the Hubbard model in $d=(1+1)$ in the high temperature
limit. The results are valid for any set of parameters that 
characterize the model. \par 

\vfill

  \par

\vskip 0.2cm

  \par

\eject

\baselineskip=18pt

\noindent {\bf 1. Introduction} \par 

A quantum system at thermal equilibrium can be completely described
provided that one knows its grand canonical partition function, which
can be expressed as a path integral. For bosonic systems, an
advantageous feature of the path integral approach is that of
employing commuting functions instead of non-commuting operators. For
fermionic systems, however, such an advantage is not obvious to hold,
as the integration variables are also non-commuting. \par

In Condensed Matter Physics we have self-interacting fermionic models
that describe correlated electrons. Certainly one of the most studied
models is the Hubbard model [1] in a variety of space dimensions. \par

Due to the non--commutative nature of the fermion fields, it is a common
practice to bosonize the fermionic model. In the path integral
formulation of the Hubbard model, this path integral can be
re--written in terms of auxiliary bosonic fields using the
Hubbard--Stratonovich transformation [2]. The Hubbard--Stratonovich
transformation allows the introduction of different decompositions
through different auxiliary bosonic fields [3]. Such decompositions
are equivalent only when the path integral can be calculated exactly.
However, due to the quartic interaction term in the Hubbard model,
only perturbative results are attained, and we can no longer relate
terms of the perturbative expansion obtained through different
decompositions [3]. The ambiguity thus arised is handled by a suitable
choice of decompositions, guided by the physical symmetries of the
problem under consideration. \par 

In the recent literature, we find many papers discussing the high
temperature expansion of the Hubbard model in space dimensions larger
than one [4]. Even though we have exact results for the Hubbard model
in one space dimension at zero temperature [5], the study of 
the behavior of the unidimensional Hubbard model in the high 
temperature limit, we only find in the work by Takahashi [6] and
 Shiba--Pincus [7] in the seventies. \par

Takahashi [6] derives  an integral equation
for the thermodynamic potential. However, closed expressions
for the physical quantities are only obtained in two limits:
$U\rightarrow 0$ and $U\rightarrow \infty$. Shiba--Pincus [7] made
a numerical analysis for the unidimensional half--filled--band
Hubbard model as a function of temperature. They considered two kinds
of boundary condition: $i$) finite chain with open ends; $ii$) finite
ring with cyclic boundary conditions. In both cases, the model studied
has at most six atoms. The results  of the finite size model 
were  used to extrapolate to the case of an infinity chain at finite
temperature.  \par

We exploited in a previous paper the grassmannian nature of the
fermionic fields to study the high temperature behavior of the grand
canonical partition function for the anharmonic fermionic
oscillator [8]. That model has zero (space) dimension, and a natural
extension is the unidimensional Hubbard model [1]. \par 

In a recent paper [9], we calculated the exact first-order coefficient
in $\beta$ ($ \beta= {1 \over kT}$) of the grand canonical
partition funtion for the unidimensional Hubbard model in the high
temperature limit. Our result was free from any ambiguity, 
like the one  introduced by the Hubbard--Stratonovich 
transformation [3]. Besides, it was valid for
any set of parameters characterizing the model. \par 

In this article, the method for evaluating the path integral of
self--interacting fermionic systems presented in references [8,9] is
improved, being thus extended to get the exact higher order 
coefficients of the grand canonical partition function for 
the unidimensional systems. \par 

Exemplifying the application of our method, we consider the
unidimensional periodic Hubbard model. The exact coefficients of
order $\beta^2$ and $\beta^3$ of the grand canonical partition
function expansion are calculated in the high temperature limit. \par 

In section 2 we present the method of evaluating multivariable
fermionic integrals, based on the grassmannian nature of the fermionic
fields. The results derived in this section are valid for any
unidimensional self--interacting fermionic model. We also write down
the kernel of the Grassmann function associated to {\bf K} in the
Hubbard model, where ${\bf K}= {\bf H}- \mu {\bf N}$ (see eq. (2.2)).
In section 3 we discuss the symmetries in one--space dimension that
allow us to show that many multivariable integrals involved in
the calculation  are equal. Those symmetries guarantee that we 
have a small number of integrals to calculate. In section 4 we 
calculate the exact coefficients of order $\beta^2$ and $\beta^3$ 
in the expansion of the grand canonical partition function in the
 high temperature limit. In section 5 we obtain the Helmholtz 
free energy of the undimensional Hubbard model up to order $\beta^3$. 
The result is valid for any number $N$ of space sites. We derive 
the average energy by site, the average diference of
number of spins up and down per site  and finally the average of 
the square of magnetization per site. In Appendix A, we show that
 the moments of gaussian Grassmann multivariable integrals are 
equal to co--factors of an appropriate matrix. \par 

\vskip 1cm 

\noindent {\bf 2. Expansion in the High Temperature Limit and the
Grassmann Multivariable Integrals} \par 

We have recently shown that the grand canonical partition function of
the Hubbard model in $d=(1+1)$, up to first order in $\beta$, for a
finite number $N$ of space sites is [9], \par 

$$\displaylines{ {\cal Z}( \beta, \mu) = \int \prod_{l=1}^{\rm N}
\prod_{\sigma= \pm 1} \prod_{i=0}^{\rm M-1} d \psi_\sigma(x_l, \tau_i)
d\bar\psi_\sigma( x_l, \tau_i) \hfill \cr 
\hskip 0.7cm 
e^{\sum\limits_{l=1}^{\rm N} \sum\limits_{\sigma=\pm 1}
\sum\limits_{i=0}^{\rm M-1} \, \bar\psi_\sigma( x_l, \tau_i) (
\psi_\sigma( x_l, \tau_i) - \psi_\sigma( x_l, \tau_{i+1}))} \; \;
e^{ - \varepsilon \sum\limits_{i=0}^{\rm M-1} {\bf K}(
\bar\psi_\sigma( x_l, \tau_i), \psi_\sigma( x_l, \tau_i))} +
O(\beta^2), \hfill (2.1) \cr } $$ 

\noindent where $\beta = {1\over kT}$, $k$ is the Boltzmann constant,
and $T$ is the absolute temperature. The space lattice has $N$ sites
and $M$ is the number of sites in the temperature lattice, such that
$\varepsilon M = \beta$. {\bf K} is given by \par 

$$ {\bf K} = {\bf H} - \mu {\bf N}, \eqno (2.2) $$ 

\noindent {\bf H} is the hamiltonian of the system, $\mu$ is the
chemical potential and {\bf N} is the total number of particles
operator. \par 

Using a previous result [10], we were ible, for fixed $N$ mnd $M$,
to re--write eq.(2.1) as a sum of determinants. We used the simbolic
manipulation language MAPLE V.3 to calculate the determinants for
fixed numerical values of $N$ and $M$, and get a recursive expression
 for the grand canonical partition function of the unidimensional 
Hubbard model up to order $\beta$: \par 

$$ {\cal Z}(\beta, \mu; B) = 2^{2{\rm N}} \biggl[ 1 - {\rm N}\beta \big(
({\rm E}_0 - \mu) + {U \over 4}\big)\biggr]. \eqno (2.3)$$ 

It is important to point out that the evaluation of determinants of
$NM \times NM$ matrices were in order. Due to memory and speed
limitations, such task got restricted to matrices for which $NM \leq
30$. It was also a matter of luck to obtain the expression (2.3).
\par 

Now, we present a new method to obtain the exact coefficients of the
expansion of the grand canonical partition function for any
self--interacting fermionic unidimensional model in the high
temperature limit. The fermionic models have $N$ space sites, where N
is any integer value. From the results derived we can calculate the
thermodynamic limit of $N\rightarrow \infty$. \par 

The grand canonical partition function in the limit of high
temperature is, \par 

\vskip -0.5cm

$$ \eqalignno{ {\cal Z}(\beta, \mu) &= {\rm Tr}( e^{- \beta {\bf K}})
& \cr 
&= {\rm Tr}[ 1\!\hskip -1pt{\rm I} - \beta {\bf K}] +
{\beta^2 \over 2} {\rm Tr}[ {\bf K}^2] - {\beta^3 \over 3!} {\rm Tr}[
{\bf K}^3]  + \cdots, & (2.4) \cr }$$ 

\noindent where {\bf K} is given by (2.2). Really,(the
second line on the r.h.s. of eq.(2.4) is the expansion of the
exponential of an operator. \par 

The trace of any fermionic operator, written in terms of creation
(${\bf a}_i^\dagger$) and destruction (${\bf a}_j$) operators, can be
mapped in terms of the generators of the Grassmann algebra $\{
\bar{\eta}_i , \eta_j \}$, provided we make the identification: \par 

$$ {\bf a}_i^\dagger \rightarrow \bar{\eta}_i \hskip 1cm {\rm and}
\hskip 1cm {\bf a}_j \rightarrow {\partial \over \partial
\bar{\eta}_j}. \eqno(2.5) $$ 

Let us consider a Grassmann algebra of dimension $2^{2{\cal N}}$, whose
generators are:
$\{ \bar{\eta}_1, \cdots , \break
\bar{\eta}_{\cal N}; \eta_1, \cdots, \eta_{\cal N} \}$. 
The generators $\eta_i, \bar{\eta}_j$ satisfy anti--commutation
 relations: \par 

$$\{{\eta}_i , \eta_j \} = 0, \hskip 0.5cm \{ \bar{\eta}_i ,\bar\eta_j
\} =0 \hskip 0.5cm {\rm and} \hskip 0.5cm \{ \bar{\eta}_i , \eta_j
\}=0, $$ 

\noindent where $i,j= 1,2, \cdots, {\cal N}$. \par 

The product of two fermionic operators {\bf A} and {\bf B}, written in
terms of the Grassmann generators, is \par 

$$ ({\bf AB}) (\bar{\eta}, \eta) = \int \prod_{I=1}^{\cal N}
d\eta_I^\prime \hskip 2pt d\bar{\eta}_I^\prime \, {\cal A}(\bar{\eta},
\eta^\prime)\, {\cal B}(\bar{\eta}^\prime, \eta) 
% 
% expressao da exponencial da soma 
%
\hskip 0.2cm e^{- \sum\limits_{j=1}^{\cal N}
\bar{\eta}_j \eta_j}, \eqno (2.6) $$ 

\noindent where ${\cal A}(\bar{\eta}, \eta^\prime)$ and ${\cal
B}(\bar{\eta}^\prime, \eta)$ are the kernel of the fermionic operators
{\bf A} and {\bf B}, respectively. \par 

For fermionic operators in normal order, \par 

$$ \eqalignno{ {\bf A} &= \sum_{n_1, \cdots, n_{\cal N} = 0 \atop{m_1,
\cdots, m_{\cal N}=0 }}^1 A_{n_1, \cdots, n_{\cal N}; m_1, \cdots,
m_{\cal N}}^{\normal} \; ({\bf a}_1^\dagger)^{n_1} \cdots ({\bf
a}_{\cal N}^\dagger)^{n_{\cal N}} \; ({\bf a}_{\cal N})^{m_{\cal N}}
\cdots ({\bf a}_1)^{m_1}, & (2.7a) \cr
\noalign{\hbox {and}}
{\bf B} &= \sum_{n_1, \cdots, n_{\cal N} =0 \atop{m_1, \cdots, m_{\cal
N} =0}}^1 B_{n_1, \cdots, n_{\cal N}; m_1, \cdots, m_{\cal
N}}^{\normal} \; ({\bf a}_1^\dagger)^{n_1} \cdots ({\bf a}_{\cal
N}^\dagger)^{n_{\cal N}} \; ({\bf a}_{\cal N})^{m_{\cal N}} \cdots
({\bf a}_1)^{m_1}, & (2.7b) \cr }$$ 

\noindent with $A_{n_1, \cdots, n_{\cal N}; m_1, \cdots, m_{\cal
N}}^{\normal}$ and $B_{n_1, \cdots, n_{\cal N}; m_1, \cdots, m_{\cal
N}}^{\normal}$ being commuting constants. The kernel of these
operators are given by, \par 

$$\eqalignno{ {\cal A}(\bar{\eta}, \eta^\prime) &= e^{\;
\sum\limits_{l=1}^{\cal N} \bar{\eta}_l \eta_l^\prime} \; {\cal
A}^{\normal}(\bar{\eta}, \eta^\prime) & (2.7c) \cr
\noalign{\hbox{and}}
{\cal B}(\bar{\eta}^\prime, \eta) &= e^{\;
\sum\limits_{l=1}^{\cal N} \bar{\eta}_l^\prime \eta_l} \;
 {\cal B}^{\normal}(\bar{\eta}^\prime, \eta) & (2.7d) \cr } $$ 

\noindent where, in a naive way, we can say that we get ${\cal
A}^{\normal}(\bar{\eta}, \eta^\prime)$ from eq.(2.7a) by replacing
${\bf a}_i^\dagger \rightarrow \bar{\eta}_i$ and ${\bf a}_j
\rightarrow \eta_j$, and similarly for ${\cal
B}^{\normal}(\bar{\eta}^\prime, \eta)$. \par 

All the operators considered in this paper are in normal order. \par 

The trace of any fermionic operator {\bf O} is [11] \par 

$$ {\rm Tr} [{\bf O}] = \int \prod_{l=1}^{\cal N} \, d\eta_i
d\bar{\eta}_i \; {\cal O}(\bar{\eta}, \eta) \; \; e^{
\;\sum\limits_{j=1}^{\cal N} \bar{\eta}_j \eta_j}, \eqno(2.8) $$ 

\noindent where we use the shorthand notation: $ \bar{\eta} \equiv \{
\bar{\eta}_1, \cdots, \bar{\eta}_{\cal N}\}$ and $\eta\equiv \{
\eta_1, \cdots,\eta_{\cal N}\}$, and ${\cal O}(\bar{\eta}, \eta)$ is
the kernel of the fermionic operator {\bf O}. For the case where the
operator {\bf O} is a product of $n$ normal ordered fermionic
operators {\bf Q}, we use the relations (2.6--8) and algebraic
manipulations to write the trace of such product of operators as \par 

$$\eqalignno{ {\rm Tr} [{\bf Q}^n] &= \int \prod_{l=1}^{\cal N}
\prod_{\mu=0}^{n-1} \, d\eta_l (\tau_\mu) d\bar{\eta}_l(\tau_\mu) \;
e^{\sum\limits_{l=1}^{\cal N} \sum\limits_{\nu=0}^{n-1}
\bar{\eta}_l(\tau_\nu) [\eta_l (\tau_\nu) - \eta_l (\tau_{\nu+1})]}
\hskip 0.3cm \times & \cr
& \times {\cal
O}^{\normal}(\bar{\eta}(\tau_0), \eta(\tau_0)) \, {\cal
O}^{\normal}(\bar{\eta}(\tau_1), \eta(\tau_1))\times \cdots \times
{\cal O}^{\normal}(\bar{\eta}(\tau_{n-1}), \eta(\tau_{n-1})), & (2.9)
\cr} $$ 

\noindent where we use the convention: $ \bar{\eta}(\tau_\nu) \equiv
\{ \bar{\eta}_1 (\tau_\nu), \cdots, \bar{\eta}_{\cal N} (\tau_\nu)\}$
and $\eta (\tau_\nu) \equiv \{ \eta_1 (\tau_\nu) , \cdots,\eta_{\cal
N} (\tau_\nu)\}$ for $\nu = 0,1, \cdots, n-1$. The Grassmann variables
in eq. (2.9) satisfy the boundary condition: \par 

$$ \eta_l (\tau_n) = - \eta_l (\tau_0) \hskip 0.5cm {\rm and} \hskip
0.5cm \eta_l (\tau_\nu) = 0, \hskip 0.7cm {\rm for} \;\;\; \nu> n,
\eqno(2.9a) $$ 

\noindent and $l=1, 2, \cdots, {\cal N}$. 

Relation (2.9) is used to write the terms of the expansion of the
grand canonical partition function in the high temperature limit as
multivariable Grassmann integrals. \par 

For a one--dimensional fermionic model, the index $l$ of the
generators of the Grassmann algebra: $\{\eta_l (\tau_\mu)$, 
$\bar{\eta}_l (\tau_\mu)\}$,  conveys space and spin degrees of freedom.
Explicity separating the space index from the spin index in the
generators of the non--commuting algebra, the coefficients of the
expansion ${\cal Z}(\beta, \mu)$ in eq.(2.4) become, \par 

$$\eqalignno{ {\rm Tr} [{\bf K}^n] &= \int \prod_{l=1}^ N
\prod_{\sigma= \pm 1} \prod_{\nu=0}^{n-1} \, d\eta_\sigma (x_l,
\tau_\nu) d\bar{\eta}_\sigma(x_l,\tau_\nu) \; e^{\sum\limits_{l=1}^ N
\sum\limits_{\sigma= \pm 1} \sum\limits_{\nu=0}^{n-1}
\bar{\eta}_\sigma(x_l, \tau_\nu) [\eta_\sigma (x_l, \tau_\nu) -
\eta_\sigma (x_l, \tau_{\nu+1})]} \hskip 0.3cm \times & \cr
&
\hskip -0.5cm \times {\cal K}^{\normal}(\bar{\eta}_\sigma(x,\tau_0),
\eta_\sigma(x, \tau_0)) \, {\cal K}^{\normal}(\bar{\eta}_\sigma(x,
\tau_1), \eta_\sigma (x, \tau_1))\times \cdots \times {\cal
K}^{\normal}(\bar{\eta}_\sigma(x, \tau_{n-1}), \eta_\sigma(x,
\tau_{n-1})), & \cr & & (2.10)\cr } $$ 

\noindent where $N$ is the number of space site, $\sigma = \uparrow=
+1, \sigma = \downarrow= -1$, and the boundary conditions (2.9a)
become, \par 

$$ \eta_\sigma(x_l, \tau_n) = - \eta_\sigma (x_l, \tau_0) \hskip 0.5cm
{\rm and} \hskip 0.5cm \eta_\sigma (x_l, \tau_\nu) = 0, \hskip 0.3cm
{\rm for}\;\; \nu> n, \eqno(2.11) $$ 

\noindent $l = 1, 2, \cdots, N$. \par 

To make use of previous results, where we relate the multivariable
Grassmann integrals to a sum of determinants [10], it is interesting
to map the generators $\eta_\sigma (x_l, \tau_\nu)$ and $\bar{
\eta}_\sigma (x_l, \tau_\nu)$ into single--indexed anti--commuting
variables. The particular mapping is a matter of choice; however,
calculations are greatly simplified if we choose: \par 

$$\eqalignno{ \eta_\uparrow (x_l, \tau_\nu) &\equiv \eta_{\nu N + l} &
(2.12a) \cr
\noalign {\hbox{ and}}
 \eta_\downarrow (x_l, \tau_\nu)
&\equiv \eta_{(n+\nu) N + l}, & (2.12b) \cr }$$ 

\noindent where $ l= 1, 2, \cdots, N$. $N$ is the number of space
sites, and, $ \nu= 0, 1, \cdots, n-1$. \par 

The mappings (2.12 a--b) can be summarized as: \par 

$$ \eta_\sigma (x_l, \tau_\nu) \equiv \eta_{[{(1-\sigma)\over 2} n +
\nu]N + l}\hskip 4pt. \eqno (2.12c) $$ 

Using the redefined generators (2.12c), the sum in the exponential on
the r.h.s. of eq.(2.10) is written as \par 

$$ \eqalignno{ \sum_{l=1}^ N &\sum_{\sigma= \pm 1} \sum_{\nu=0}^{n-1}
\bar{\eta}_\sigma(x_l, \tau_\nu) [\eta_\sigma (x_l, \tau_\nu) -
\eta_\sigma (x_l, \tau_{\nu+1})]= & \cr
& = \sum_{I, J=1}^{2nN}
\bar{\eta}_I \;A_{ I J}\; \eta_J, & (2.13) \cr }$$ 

\noindent where $A_{I J}$ are the entries of the block--matrix {\bf
A}, \par 

$$ {\bf A} = \pmatrix { {\bf E}^{\uparrow \uparrow} & \0barra \cr
 & &   \cr
   \0barra & {\bf A}^{\downarrow \downarrow} \cr }, \eqno (2.13a) $$ 

\noindent whose entries are matrices of dimension $nN\times nN$. The
indices $I,J$ are such that $I,J = 1, 2, \cdots, nN$. \par 

The matrices ${\bf A}^{\uparrow \uparrow}$ and ${\bf A}^{\downarrow
\downarrow}$ are identical and have a block--structure. Taking into
account the anti--periodic condition in temperature (2.11) in (2.13),
the matrices ${\bf A}^{\sigma \sigma}$, $\sigma=\uparrow, \downarrow$,
are \par 

$$ {\bf A}^{\uparrow \uparrow} = {\bf A}^{\downarrow \downarrow} =
\pmatrix { \lbarra_{N\times N} & - \lbarra_{N\times N} &
\0barra_{N\times N} & \cdots & \0barra_{N\times N} \cr & & & & \cr
%
%segunda linha da matriz
%
\0barra_{N\times N} & \lbarra_{N\times N}&
- \lbarra_{N\times N} & \cdots & \0barra_{N\times N}\cr
%
%terceira linha da matriz
%
& & & & \cr \vdots & & & & \vdots \cr
%
% ultima linha da matriz
%
 \lbarra_{N\times N} & \0barra_{N\times N} &
\0barra_{N\times N} & \cdots & \lbarra_{N\times N} \cr}. \eqno(2.14a)
$$ 

The non--null elements of ${\bf A}^{\sigma \sigma}$, $\sigma=\uparrow$
and $\sigma= \downarrow$, are: \par 

$$ A_{I J}^{\sigma \sigma} = \cases{ a_{II} = 1, & $I= 1, 2, \cdots,
nN $ \cr & \cr a_{I, I+N} = -1, & $I= 1, 2, \cdots, (n-2)N $ \cr & \cr
a_{(n-1)N +I, I} = 1, & $ I= 1, 2, \cdots, N $ \cr} \eqno(2.14b) $$ 

\vskip 0.5cm 

The matrices ${\bf A}^{\sigma \sigma}$, $\sigma = \uparrow,
\downarrow$, have dimension $nN\times nN$. The symbols
$\lbarra_{N\times N}$ and $\0barra_{N\times N}$ stand for the identity
and zero matrices of dimension $N\times N$, respectively.

With the newly indexed generators, the expression of ${\rm Tr}[{\bf
K}^n]$ (eq.(2.10)) becomes, \par 

$$\eqalignno{ {\rm Tr}[{\bf K}^n] &= \int \prod_{I=1}^{2nN} \, d\eta_I
d\bar{\eta}_I \;\; e^{\sum\limits_{I,J= 1}^{2nN} \bar{\eta}_I\; A_{I
J} \; \eta_J} \times & \cr
%
%segunda linha
%
 & \hskip -1cm \times
{\cal K}^{\normal} (\bar{\eta}, \eta; \nu=0)\; {\cal K}^{\normal}
(\bar{\eta}, \eta; \nu=1) \; \cdots \; {\cal K}^{\normal} (\bar{\eta},
\eta; \nu=n-1). & (2.15) }$$

Note that expression (2.15) is the exact coefficient in order
$\beta^n$ of the expansion in the high-temperature limit of the grand
canonical partition function for any unidimensional self--interacting
fermionic model. The specific model to be studied is represented by
the function ${\cal K}^{\normal}$. The matrix {\bf A} is the same for
all unidimensional fermionic models. \par

The Grassmann functions ${\cal K}^{\normal}$ are polynomials in the
generators of the algebra. Therefore, the r.h.s. of eq.(2.15) are
moments of the multivariable Grassmann gaussian integrals. We show in
Appendix A that these integrals can be written as co--factors of the
matrix {\bf A}. \par 

Once the sub--matrices ${\bf A}^{\uparrow \downarrow}$ and ${\bf
A}^{\downarrow \uparrow}$ are null, the multivariable integral (2.15)
is equal to the product of the contributions coming from the sectors:
$ \sigma \sigma= \uparrow \uparrow$ and $ \sigma \sigma= \downarrow
\downarrow$. \par 

Even restricting ourselves to the calculation of the multivariable
integrals of a fixed sector $\sigma\sigma$, we have to obtain the
determinant of non--diagonal matrices of dimension $nN\times nN$.
The evaluation of such determinants by means of some symbolic
manipulation language obviously depends on hardware and software
resources. For a fixed $n$, for instance, we have an upper practical
limit for $N$ for doing the calculation of the determinants.
 One possibility for evaluating (2.15) is to fix
different values for $N$ and try to extrapolate the results for an
arbitrary value of $N$. If we are lucky, we may recognize some
recursion expression for (2.15) for all $N$. \par 

The integrals in (2.15), for each sector $\sigma\sigma$, have the
form: 

$$M(L,K) = \int \prod_{i = 1}^{\rm nN} d\eta_i d\bar\eta_i
 \;
\bar\eta_{l_1} \eta_{k_1} \cdots \bar\eta_{l_m} \eta_{k_m} \hskip 3pt
 e^{ \sum\limits_{ i, j = 1}^{\rm nN} \bar\eta_i A_{i j} \eta_j},
\eqno(2.16) $$ 

\noindent with $ L= \{ l_1, \cdots, l_m\}$ and $K= \{k_1, \cdots,
k_m\}$. The products $\bar{\eta}\eta$ are ordered in such a way that
$ l_1< l_2< \cdots< l_m$ and $ k_1< k_2< \cdots< k_m$. From eq.(A.10),
the result of this type of integrals is equal to: \par 

$$ M(L,K) = (-1)^{(l_1 + l_2+ \cdots+ \l_m) + (k_1+ k_2+ \cdots+ k_m)}
A(L,K), \eqno(2.17) $$ 

\noindent where $A(L,K)$ is the determinant of the matrix obtained
from matrix {\bf A} by deleting the lines $\{ l_1, \cdots, l_m\}$ and
the columns: $\{k_1, \cdots, k_m\}$. \par 

Our approach to calculate the integral (2.16) is, for fixed $n$ and
arbritary $N$, to explore the block--structure of matrices ${\bf
A}^{\sigma \sigma}$, $\sigma=\uparrow$ and $\sigma=\downarrow$,
diagonalizing it through a similarity transformation \par 

$$ {\bf P}^{-1} {\bf A}^{\sigma\sigma} {\bf P} = {\bf D}, \eqno(2.18a)
$$ 

\noindent where the matrix {\bf D} is, \par 

$$ {\bf D} = \pmatrix{ \lambda_1 \lbarra_{N\times N} &
\0barra_{N\times N} & \cdots & \0barra{N\times N} \cr
\0barra_{N\times N} & \lambda_2 \lbarra_{N\times N} & \cdots &
\0barra_{N\times N} \cr
 \vdots & & & \vdots\cr
\0barra_{N\times N} & \0barra_{N\times N} & \cdots & \lambda_n
\lbarra_{N\times N} \cr }, \eqno(2.18b)$$ 

\noindent $\lambda_i$, $i= 1, 2, \cdots, n$, are the eigenvalues of
matrices ${\bf A}^{\sigma\sigma}$, $\sigma= \uparrow, \downarrow$, and
calculate the co--factors of the matrix {\bf D}. \par 

The $j^{th}$--column of matrix {\bf P} is the eigenvector of ${\bf
A}^{\sigma\sigma}$ associated to the eigenvalue $\lambda_j$. Each
eigenvalue of matrix ${\bf A}^{\sigma\sigma}$ has degeneracy $N$. The
matrices are not hermitian, thus some eigenvalues are complex. In
the following, we will be working on the
$\sigma\sigma=\uparrow\uparrow$ sector; however, the results for the
$\sigma\sigma=\downarrow\downarrow$ sector are analogous since ${\bf
A}^{\uparrow\uparrow} = {\bf A}^{ \downarrow\downarrow}$. \par 

We make a change of anti--commuting variables, \par 

$$ \eta^\prime = {\bf P}^{-1} \, \eta \hskip 1cm {\rm and} \hskip 1cm
\bar{\eta}^\prime = \bar{\eta} \,{\bf P}, \eqno(2.19) $$ 

\noindent where $\eta^\prime \equiv \{ \eta_1^\prime ,
\cdots,\eta_{nN}^\prime \}$ and $ \bar{\eta}^\prime \equiv \{
\bar{\eta}_1^\prime, \cdots, \bar{\eta}_{nN}^\prime \}$. The jacobian
of the transformation (2.19) is equal to one. \par 

In eqs.(2.19), we wrote the transformations of variables in a very
simplified way; however, due to the fact that ${\bf
A}^{\uparrow\uparrow}$ is a block--matrix, the matrix {\bf P} also has
a block structure. This fact implies that transformations (2.19)
do not mix the index associated to the space site. \par 

In a schemmtic way, the integrals M(L,K) (eq.(2.16)) become: \par 

$$ \eqalignno{ M(L,K) &= \int \prod_{i = 1}^{\rm nN} d\eta_i
d\bar\eta_i
 \; (\bar\eta {\bf P}^{-1})_{l_1} ({\bf P}\eta)_{k_1}
\cdots (\bar\eta {\bf P}^{-1})_{l_m} ({\bf P} \eta)_{k_m} 
	\hskip 3pt & \cr
 & \hskip 3cm e^{ \sum\limits_{ i, j = 1}^{\rm nN} \bar\eta_i D_{i
j} \eta_j}, & (2.20) \cr }$$ 

\noindent where $D_{i j}$ are the entries of matrix {\bf D}. \par 

From eq.(2.16), we have that $M(L,K)$ is a multivariable Grassmann
 integral, which in turn correspond to co--factors of the 
diagonalized matrix {\bf D} (eq.(2.17)). It is very simple to
 calculate these co--factors. Besides, taking into
account that $\eta^{\prime^2} =0$ and $ \bar{\eta}^{\prime^2} =0$
decreases the number of integrals to be calculated in (2.20). \par 

This is a general approach, and it can be applied to any
self--interacting unidimensional fermionic model. Thus we have reduced
the calculation of fermionic path integral to the evaluation of
co--factors of a diagonalized matrix. \par  

\vskip 1cm 

\noindent {\bf 2.1. Application to Unidimensional Periodic Hubbard Model} 

\bigskip 

We will apply in section 4 the general approach presented in section 2
to the periodic Hubbard model [1] in $d=(1+1)$ with $N$ space sites
in the presence of an external magnetic field in the $\hat{z}$
direction. However, we need to know the normal ordered Grassmann
function ${\cal K}^{\normal}$ for the specific model. \par 

The hamiltonian that describes the Hubbard model in one space
dimension is [1]: \par 

$${\bf H} = \sum_{ {i=1}\atop{j=1}}^{\rm N} \sum_{\sigma= -1, 1}
t_{ij} {\bf a}_{i\sigma}^{\dagger}{\bf a}_{j\sigma} + U \sum_
{i=1}^{\rm N} {\bf a}_{i\uparrow}^\dagger{\bf a}_{i\uparrow} {\bf
a}_{i\downarrow}^\dagger{\bf a}_{i\downarrow} + \lambda_B \sum_
{i=1}^{\rm N} \sum_{\sigma=-1, 1} \sigma {\bf a}_{i\sigma}^\dagger{\bf
a}_{i\sigma} \eqno (2.1.1) $$ 

\noindent where ${\bf a}_{i\sigma}^\dagger$ is the creation operator
of an electron in site {\it i} with spin [12]  $\sigma$ and
 ${\bf a}_{i\sigma}$  is the destruction
operator of an electron in site {\it i} with spin $\sigma$. All
diagonal elements of $t_{ij}$ are equal, $t_{ii} = {\rm E_0}$, where
E$_0$ is the kinetic energy. The only non--null off--diagonal terms
are $t_{i, i-1}=t_{i, i+1}= t$, where $ i=1, 2, \dots , {\rm N}$, and
they contribute to the hopping term. {\it U} is the strenght of the
interaction between the electrons in the same site but with different
spins. We have defined $\lambda_B= -{1\over 2} g\mu_B B$, where {\it
g} is the Land\'e's factor and $\mu_B$ is the Bohr's magneton. \par 

The periodic boundary condition in space is implemented by imposing
that ${\bf a}_{0 \sigma} \equiv {\bf a}_{N \sigma}$ and ${\bf a}_{N+1,
\sigma} \equiv {\bf a}_{1 \sigma}$. Therefore, the hopping terms $t_{1
0} {\bf a}_{1 \sigma}^\dagger {\bf a}_{0 \sigma}$ and $t_{N, N+1} {\bf
a}_{N \sigma}^\dagger {\bf a}_{N+1, \sigma}$ become $t_{1 N} {\bf
a}_{1 \sigma}^\dagger {\bf a}_{N\sigma}$ and $t_{N, 1} {\bf a}_{N
\sigma}^\dagger {\bf a}_{1 \sigma}$ respectively. We point out that
the hamiltonian (2.1.1) is already in normal order. \par 

From eq.(2.2) we have \par 

$$ {\bf K} = {\bf H} - \mu {\bf N}. $$ 

The kernel of the operator {\bf K} of a one--dimensional Hubbard model
in a lattice with $N$ space sites, written in terms of the generators
$\eta_\sigma (x_l, \tau_\nu)$ and $\bar{\eta}_\sigma (x_l, \tau_\nu)$
is \par 

$$\eqalignno{ {\cal K}^{\normal} &(\bar{\eta}_\sigma (x_l, \tau_\nu),
\eta_\sigma (x_l, \tau_\nu)) = \sum_{l=1}^N \sum_{\sigma\pm 1} (E_0 +
\sigma \lambda_B - \mu) \bar{\eta}_\sigma (x_l, \tau_\nu)\eta_\sigma
(x_l, \tau_\nu) + & \cr
%
% segunda linha da expressao
%
 & +
\sum_{l=1}^N \sum_{\sigma\pm 1} t [ \bar{\eta}\sigma (x_l,
\tau_\nu)\eta_\sigma (x_{l+1}, \tau_\nu) + \bar{\eta}_\sigma (x_l,
\tau_\nu)\eta_\sigma (x_{l-1}, \tau_\nu)] + & \cr
%
% terceira linha da expressao
%
 & + \sum_{l=1}^N U \bar{\eta}_\uparrow (x_l,
\tau_\nu)\eta_\uparrow (x_l, \tau_\nu) \bar{\eta}_\downarrow (x_l,
\tau_\nu)\eta_\downarrow (x_l, \tau_\nu), & (2.1.2) \cr }$$ 

\noindent with the space periodic boundary condition imposed by: \par 

$$ \eqalignno{ t \bar{\eta}_\sigma (x_N, \tau_\nu) \eta_\sigma
(x_{N+1}, \tau_\nu) & \equiv t \bar{\eta}_\sigma (x_N, \tau_\nu)
\eta_\sigma (x_1, \tau_\nu)  & (2.1.2a)  \cr
\noalign{\hbox{and}}
t \bar{\eta}_\sigma (x_1, \tau_\nu) \eta_\sigma (x_0, \tau_\nu) & \equiv
t \bar{\eta}_\sigma (x_1, \tau_\nu) \eta_\sigma (x_N, \tau_\nu), &
(2.1.2b) \cr } $$ 

\noindent and the anti--periodic boundary condition in $\tau$: \par 

$$ \eta_\sigma (x_l, \tau_n) = - \eta_\sigma (x_l, \tau_0), \eqno
(2.1.2c) $$ 

\noindent for $l= 1, 2, \cdots, N$, and $\sigma= \uparrow,
\downarrow$. \par 

Using the mapping (2.12c), the kernel of the operator {\bf K} becomes:
\par 

$$\eqalignno{ {\cal K}^{\normal} &(\bar{\eta},\eta;\nu)) =
\sum_{l=1}^N \sum_{\sigma\pm 1} (E_0 + \sigma \lambda_B - \mu) \;
\bar{\eta}_{[{(1-\sigma)\over 2}n +\nu]N + l}\;
\eta_{[{(1-\sigma)\over 2}n +\nu]N + l} + & \cr
%
% segunda linha da expressao
%
& + \sum_{l=1}^N \sum_{\sigma\pm 1} t [
\bar{\eta}_{[{(1-\sigma)\over 2}n +\nu]N + l}\;
\eta_{[{(1-\sigma)\over 2}n +\nu]N + l+1}+
\bar{\eta}_{[{(1-\sigma)\over 2}n +\nu]N + l}\;
\eta_{[{(1-\sigma)\over 2}n +\nu]N + l-1}] + & \cr
%
%terceira linha da expressao
%
 & + \sum_{l=1}^N U \bar{\eta}_{(n+\nu)N + l}\;\;
\eta_{(n+\nu)N + l}\;\; \bar{\eta}_{\nu N + l} \; \; \eta_{\nu N + l}.
& (2.1.3) \cr }$$ 

\vskip 1cm 

%\vfill
%\eject

\noindent {\bf 3. Useful Symmetries of the Multivariable Grassmann
Integrals} 

\bigskip 

In eq.(2.15), the number of terms to be calculated increases rapidly
with $N$. To make this approach feasible, we have to explore
symmetries, besides the trivial space translation, of the exponential
in eq.(2.15). In section 2, we pointed out that in expression (2.15) the
contributions from the sectors $\sigma\sigma= \uparrow\uparrow$ and
$\sigma\sigma= \downarrow\downarrow$ are decoupled. In this section we
discuss only one of the two sectors, e.g. $\sigma\sigma=
\uparrow\uparrow$. Analogous results are valid for the sector
$\sigma\sigma= \downarrow\downarrow$. In the expression of 
${\rm Tr} [{\bf K}^n]$ (eq.(2.15)), we deal with integrals of the
 following type: \par 

$$ \eqalignno{ I(m) &= \int \prod_{I=1}^{nN} \, d\eta_I d\bar{\eta}_I
\; \bar{\eta}_{\nu_1 N +l_1} \eta_{\nu_1 N + k_1} \, \bar{\eta}_{\nu_2
N +l_2} \eta_{\nu_2 N + k_2} \times \cdots \times & \cr
%
% segunda linha
%
 & \times \, \bar{\eta}_{\nu_m N +l_m} \eta_{\nu_m N + k_m}
\times e^{ \sum\limits_{I,J=1}^{nN} \bar{\eta}_I \;A_{IJ}^{\uparrow
\uparrow} \; \eta_J}, & (3.1) \cr }$$ 

\noindent where $m \leq n$, and, $\nu_1 \leq \nu_2 \leq \cdots \leq
\nu_m$. \par 

We want to study how the integral $I(m)$ transforms under a change of
variables in the temperature parameter $\tau$. Consider the change of
variables: \par 

$$\eqalignno{ \eta_{\nu N + l}^\prime &= \eta_{(\nu - 1)N + l}, \hskip
1cm \nu= 1, 2, \cdots, n-1, & (3.2a) \cr
%
% segunda linha
%
\noalign{\hbox{and}} \eta_l^\prime &= - \eta_{(n-1)N + l}; & (3.2b)
\cr
%
% linha vazia
%
& & \cr
%
% terceira linha
%
 \bar{\eta}_{\nu N +
l}^\prime &= \bar{\eta}_{(\nu - 1)N + l}, \hskip 1cm \nu= 1, 2,
\cdots, n-1, & (3.2c) \cr
%
% quarta linha
%
\noalign{\hbox{and}}
\bar{\eta}_l^\prime &= - \bar{\eta}_{(n-1)N + l}. & (3.2d) \cr }$$ 

In this relabelling of generators, we are shifting the temperature
index by one unit, except for $\eta_{n-1}$ and $\bar{\eta}_{n-1}$,
identified to $-\eta_{0}$ and $-\bar{\eta}_{0}$, respectively. The
jacobian of the simultaneous transformations (3.2) is equal to one.
\par 

Before discussing how the integral $I(m)$ transforms under (3.2),
let us show that the sum in the exponentiml of eq.(3.1) i{ invarimnt
under such transformations. \par 

From eq.(2.14b), we have that the non--null terms of ${\bf
A}^{\uparrow\uparrow}$ are: \par 

$$ A_{I J}^{\sigma \sigma} = \cases{ a_{II} = 1, & $I= 1, 2, \cdots,
nN $ \cr & \cr a_{I, I+N} = -1, & $I= 1, 2, \cdots, (n-2)N $ \cr & \cr
a_{(n-1)N +I, I} = 1, & $ I= 1, 2, \cdots, N $ \cr} \eqno(3.3) $$ 

\vskip 0.5cm 

Therefore, the sum within the exponential in eq.(3.1) can be written
as, \par 

$$\eqalignno{ \sum\limits_{I,J=1}^{nN} \bar{\eta}_I \;
A_{IJ}^{\uparrow\uparrow} \; \eta_J &= \sum\limits_{\nu=0}^{n-2}
\sum\limits_{l=1}^{N} \bar{\eta}_{\nu N+l} \;\; A_{\nu N+l, \nu
N+l}^{\uparrow\uparrow} \;\; \eta_{\nu N+l} + & \cr
%
% segunda linha
%
& + \sum\limits_{l=1}^N \bar{\eta}_{(n-1) N+l} \;\; A_{(n-1)N+l,
(n-1)N+l}^{\uparrow\uparrow}\;\; \eta_{(n-1)N+l} + & \cr
%
% terceira linha
&  + \sum\limits_{\nu=0}^{n-3} \sum\limits_{l=1}^N
\bar{\eta}_{\nu N+l} \; \;A_{\nu N+l, (\nu+1)N+l}^{\uparrow\uparrow}
\;\; \eta_{(\nu+1)N+l} + & \cr
%
% quarta linha
%
 & + \sum\limits_{l=1}^{N} \bar{\eta}_{(\nu-2)N+l} \; A_{(\nu-2)N+l,
(\nu-1)N+l}^{\uparrow\uparrow} \eta_{(\nu-1)N+l} + & \cr
%
% quinta linha
%
 & + \sum\limits_{l=1}^{N} \bar{\eta}_{(n-1)N+l} \;\;
A_{(n-1)N+l, l}^{\uparrow\uparrow} \;\; \eta_{l}. & (3.4) \cr }$$ 

Under the change of variables (3.2), the terms on the r.h.s. of (3.4)
transforms as: \par 

$$\eqalignno{ \hskip -2cm 1) \sum\limits_{\nu=0}^{n-2}
\sum\limits_{l=1}^N \bar{\eta}_{\nu N+l} \;\; A_{\nu N+l, \nu
N+l}^{\uparrow\uparrow} \eta_{\nu N+l} &= & \cr
%
% segunda linha
%
& \hskip -2cm = \sum\limits_{\nu=0}^{n-2} \sum\limits_{l=1}^N \;
\bar{\eta}_{\nu N+l}^\prime \;\; A_{(\nu-1)N+l,
(\nu-1)N+l}^{\uparrow\uparrow} \;\; \eta_{\nu N+l}^\prime \cr
%
% terceira linha
%
 & \hskip -2cm = \sum\limits_{\nu=1}^{n-1}
\sum\limits_{l=1}^N \; \bar{\eta}_{\nu N+l}^\prime \;\; A_{\nu N+l,
\nu N+l}^{\uparrow\uparrow} \;\; \eta_{\nu N+l}^\prime. & (3.4a) \cr
}$$ 

$$\eqalignno{
 \hskip -5.5cm 2) \sum\limits_{l=1}^{N} \bar{\eta}_{N
(n-1)+l} \;\; A_{ N (n-1) +l, N (n-1)+l}^{\uparrow\uparrow} 
\eta_{N(n-1)+l} = & \cr
%
% segunda linha
%
& \hskip -2cm =
\sum\limits_{l=1}^N \; \bar{\eta}_{l}^\prime \;\; A_{l,
l}^{\uparrow\uparrow} \;\; \eta_l^\prime. & (3.4b) \cr }$$ 

$$\eqalignno{ \hskip -1cm 3) \sum\limits_{\nu=0}^{n-3}
\sum\limits_{l=1}^N \; \bar{\eta}_{\nu N+l} \;\; A_{\nu N+l,
(\nu+1)N+l}^{\uparrow\uparrow} \;\; \eta_{(\nu+1)N+l} &= & \cr
%
% segunda linha
%
& \hskip -3cm = \sum\limits_{\nu=1}^{n-2}
\sum\limits_{l=1}^N \bar{\eta}_{\nu N+l}^\prime \;\; A_{(\nu-1)N+l,
\nu N+l}^{\uparrow\uparrow} \;\; \eta_{(\nu+1)N+l}^\prime & \cr
%
% terceira linha
%
 & \hskip -3cm = \sum\limits_{i=1}^{n-2}
\sum\limits_{l=1}^N \; \bar{\eta}_{\nu N+l}^\prime \;\; A_{\nu N+l,
(\nu+1)N+l}^{\uparrow\uparrow} \;\; \eta_{(\nu+1)N+l}^\prime. & \cr &
& (3.4c)\cr }$$ 

$$\eqalignno{ \hskip -2.5cm 4) \sum\limits_{l=1}^N \;
\bar{\eta}_{(n-2)N+l} \;\; A_{(n-2)N+l, (n-1)N+l}^{\uparrow\uparrow}
\;\; \eta_{(n-1)N+l} &= & \cr
%
%segunda linha
%
& \hskip -4cm = -
\sum\limits_{l=1}^N \; \bar{\eta}_l^\prime \;\;
A_{(n-1)N+l,l}^{\uparrow\uparrow} \;\; \eta_{N+l}^\prime & \cr
%
%terceira linha
%
 & \hskip -4cm = \sum\limits_{l=1}^N \;
\bar{\eta}_l^\prime \;\; A_{l, N+l}^{\uparrow\uparrow} \;\;
\eta_{N+l}^\prime. & (3.4d) \cr }$$ 

$$\eqalignno{ \hskip -3cm 5) \sum\limits_{l=1}^N \;
\bar{\eta}_{(n-1)N+l} \;\; A_{(n-1)N+l, l}^{\uparrow\uparrow} \;\;
\eta_l &= & \cr
%
%segunda linha
%
 & \hskip -2cm = -
\sum\limits_{l=1}^N \; \bar{\eta}_{(n-1)N+l}^\prime \;\; A_{(n-2)N+l,
(n-1)N+l}^{\uparrow\uparrow} \;\; \eta_l^\prime & \cr
%
%terceira linha
%
 & \hskip -2cm = \sum\limits_{l=1}^N \;
\bar{\eta}_{(n-1)N+l}^\prime \;\; A_{(n-1)N+l,l}^{\uparrow\uparrow}
\;\; \eta_l^\prime. & (3.4e) \cr }$$ 

To obtain the results (3.4a--e), we have used eq.(3.3). \par 

Summing the r.h.s. of (3.4a--e), we finally conclude that  \par

$$ \sum\limits_{I,J=1}^{nN} \bar{\eta}_I\;\; A_{IJ}^{\uparrow\uparrow}
\;\; \eta_J = \sum\limits_{I,J=1}^{nN} \bar{\eta}_I^\prime \;\;
A_{IJ}^{\uparrow\uparrow} \;\; \eta_J^\prime. \eqno(3.5)$$ 

We point out that the invariance of the sum (3.5) is true for any 
value of $n$ and $N$. Moreover, since 
${\bf A}^{\uparrow\uparrow} = {\bf A}^{\downarrow\downarrow}$, a
similar invariance under (3.2) can be written for ${\bf A}^
{\downarrow\downarrow}$. \par

In (3.2) we translated the temperature index by one unit. However, the
equality (3.5) is still valid if we translate the temperature index by
any integer $\nu_0$, either positive or negative, i.e., \par 

$$\eqalignno{ \eta_{\nu N+l} \hskip 0.5cm &\rightarrow \hskip 0.5cm
\eta_{(\nu+\nu_0)N+l} & (3.6a) \cr
%
%segunda linha
%
\noalign{\hbox{and}} \eta_{\bar{\nu} N+l} \hskip 0.5cm &\rightarrow
\hskip 0.5cm - \eta_l , \hskip 1cm {\rm where} \hskip 0.5cm \bar{\nu}
+ \nu_0 = n, & (3.6b) \cr }$$ 

\noindent and perform an analogous transformation for $\bar{\eta}$.
Transformation (3.6) is equivalent to applying $\nu_0$ distinct and
consecutive transformations (3.2). \par 

Under the transformation (3.6), the integral $I(m)$ (eq.(3.1))
becomes, \par 

$$ \eqalignno{ I(m) &= \int \prod_{I=1}^{nN} \, d\eta_I^\prime
d\bar{\eta}_I^\prime \; \bar{\eta}_{(\nu_1+\nu_0) N +l_1}^\prime
\eta_{(\nu_1+\nu_0) N + k_1}^\prime \, \times \cdots \times & \cr
%
%segunda linha
%
 & \times \, \bar{\eta}_{(\nu_m+\nu_0) N +l_m}^\prime
\eta_{(\nu_m+\nu_0) N + k_m}^\prime \times e^{ \sum\limits_{I,J=1}^{nN}
\bar{\eta}_I^\prime \; A_{IJ}^{\uparrow \uparrow} \;\eta_J^\prime}, &
(3.7) \cr }$$ 

\noindent where $\nu_0$ is a fixed integer, positive or negative;
for $\bar{\nu}$ such that $\bar{\nu} +\nu_0=0 $, we assume that
\par 

$$\eqalignno{ \bar{\eta}_{(\bar{\nu} +\nu_0)N + l} \rightarrow -
\bar{\eta}_l \hskip 0.5cm &{\rm and} \hskip 0.5cm \eta_{(\bar{\nu}
+\nu_0)N + l} \rightarrow - \eta_l, \cr
%
%segunda linha
%
\noalign{\hbox{ and for $\nu+\nu_0>n$}}
 \bar{\eta}_{(\nu +\nu_0)N +
l} \rightarrow \bar{\eta}_{(\nu +\nu_0-n)N + l} \hskip 0.5cm &{\rm
and} \hskip 0.5cm \eta_{(\nu +\nu_0)N + l} \rightarrow \eta_{(\nu
+\nu_0-n)N + l}. \cr }$$ 

An analogous result is valid for the sector $\sigma\sigma=
\downarrow\downarrow$. \par 

As the contributions to the integrals in eq.(2.15) of the sectors
$\sigma\sigma= \uparrow\uparrow$ and $\sigma\sigma=
\downarrow\downarrow$ factorize, we can choose differents values
for $\nu_0$ for each spin sector. \par 

Another symmetry of the exponential in eq.(3.1) results from the
arbitrariness of either clockwise or counterclockwise labelling of
space sites in the periodic unidimensional chain. The result of the
integrals is choice--independent. Let us label the space sites
counterclockwise. For a fixed $\nu$, the Grassmann generators
associated to the sites are: $ \{\eta_{\nu N +1}, \eta_{\nu N +2},
\cdots, \eta_{\nu N +N};   \break
 \bar{\eta}_{\nu N +1}, \bar{\eta}_{\nu N +2},
\cdots, \bar{\eta}_{\nu N +N} \}$. If, on the other hand, we label the
unidimensional chain clockwise, in this case the Grassmann generators
will be: 
$ \{\eta_{\nu N +1}^\prime, \eta_{\nu N +2}^\prime, \cdots,
\eta_{\nu N +N}^\prime ;  \bar{\eta}_{\nu N +1}^\prime,    \break
\bar{\eta}_{\nu N +2}^\prime, \cdots, \bar{\eta}_{\nu N +N}^\prime \}$. 
\par 

The correspondence among generators in the two labellings may be
 chosen to be   \par

$$\eqalignno{ \eta_{\nu N+1} = \eta_{\nu N+1}^\prime \hskip 0.5cm
&{\rm and} \hskip 0.5cm \bar{\eta}_{\nu N+1} = \bar{\eta}_{\nu
N+1}^\prime, & (3.8a) \cr
 \noalign{\hbox{for $\nu= 0, 1, \cdots,
n-1$, and }}
%
%segunda linha
%
\eta_{\nu N+l} = \eta_{\nu
N+(N+2-l)}^\prime \hskip 0.5cm &{\rm and} \hskip 0.5cm \bar{\eta}_{\nu
N+l} = \bar{\eta}_{\nu N+(N+2-l)}^\prime. & (3.8b) \cr }$$ 

The jacobian associated to (3.8) is equal to one. \par 

The sum within the exponential in the integrand of eq.(3.1), namely,
\par 

$$\eqalignno{ \sum\limits_{I,J=1}^{nN} \; \bar{\eta}_I \;\;
A_{IJ}^{\uparrow\uparrow} \;\; \eta_J &= \sum\limits_{\nu=0}^{n-1}
\sum\limits_{l=1}^N \; \bar{\eta}_{\nu N+l} \;\; A_{\nu N+l, \nu
N+l}^{\uparrow\uparrow} \;\; \eta_{\nu N+l} + & \cr
%
%segunda linha
%
&+ \sum\limits_{\nu=0}^{n-2} \sum\limits_{l-1}^N \; 
\bar{\eta}_{\nu N+l} \;\; A_{\nu N+l, (\nu+1)N+l}^{\uparrow\uparrow}
 \;\;
\eta_{(\nu+1)N+l} + & \cr
%
%terceira linha
%
& + \sum\limits_{l=1}^{N} \; \bar{\eta}_{(n-1)N+l}\;\;
 A_{(n-1)N+l, l}^{\uparrow\uparrow} \;\; \eta_l, & (3.9) \cr }$$ 

\noindent is invariant under the substitution (3.8). In order to
show this, we
use the relation among the non--zero elements of ${\bf
A}^{\uparrow\uparrow}$, given by eq.(3.3), and manipulate the
expressions in a similar way as we have done before. In summary, we
get that

$$ \sum\limits_{I,J=1}^{nN} \bar{\eta}_I\;\; A_{IJ}^{\uparrow\uparrow}
\;\; \eta_J = \sum\limits_{I,J=1}^{nN} \bar{\eta}_I^\prime \;\;
A_{IJ}^{\uparrow\uparrow} \;\; \eta_J^\prime, \eqno(3.10)$$ 

\noindent where the relation between $\{\eta, \bar{\eta} \}$ and
$\{\eta^\prime, \bar{\eta}^\prime \}$ is given by (3.8). \par 

The integral $I(m)$ (see eq.(3.1)) subjected to the change of variables
(3.8) becomes, \par 

$$ \eqalignno{ I(m) &= \int \prod_{I=1}^{nN} \, d\eta_I^\prime
 d\bar{\eta}_I^\prime
\; \bar{\eta}_{\nu_1 N +(N+2-l_1)}^\prime \eta_{\nu_1 N +
(N+2-k_1)}^\prime \; \times \cdots \times & \cr
%
%segunda linha
%
 & \times \, \bar{\eta}_{\nu_m N +(N+2-l_m)}^\prime 
\eta_{\nu_m N +(N+2-k_m)}^\prime 
\times e^{ \sum\limits_{I,J=1}^{nN} \bar{\eta}_I^\prime
\;A_{IJ}^{\uparrow \uparrow}\; \eta_J^\prime}. & (3.11)\cr } $$ 

The chiral space symmetry, represented by the transformations (3.8),
is valid for any positive integers integers $n$ and $N$. \par 

From eq.(2.15), we have that \par 

$$\eqalignno{ {\rm Tr}[{\bf K}^n] &= \int \prod_{I=1}^{2nN} \, d\eta_I
d\bar{\eta}_I \;\; e^{\sum\limits_{I,J= 1}^{2nN} \bar{\eta}_I\; A_{I
J} \; \eta_J} \times & \cr
%
%segunda linha
%
 & \hskip -1cm \times
{\cal K}^{\normal} (\bar{\eta}, \eta; \nu=0)\; {\cal K}^{\normal}
(\bar{\eta}, \eta; \nu=1) \; \cdots \; {\cal K}^{\normal} (\bar{\eta},
\eta; \nu=n-1), & (3.12) }$$ 

\noindent where the matrix {\bf A} is given by (2.13a). For the
one--dimensional Hubbard model with $N$ space sites, the kernel ${\cal
K}^{\normal} (\bar{\eta}, \eta; \nu)$ is given by eq.(2.1.3). \par 

In order to simplify the algebraic manipulation for calculating
eq.(3.12), we define: \par 

$$ \eqalignno{ {\cal E} (\bar{\eta}, \eta; \nu; \sigma) &\equiv
\sum\limits_{l=1}^N \bar{\eta}_{[{(1-\sigma)\over 2}n +\nu]N + l}\;
\eta_{[{(1-\sigma)\over 2}n +\nu]N + l}\;\;; & (3.13a) \cr
%
%segunda linha
%
{\cal T}^{-} (\bar{\eta}, \eta; \nu; \sigma) & \equiv
\sum\limits_{l=1}^N \bar{\eta}_{[{(1-\sigma)\over 2}n +\nu]N + l}\;
\eta_{[{(1-\sigma)\over 2}n +\nu]N + l+1} \;\; ; & (3.13b)\cr
\noalign{\hbox{and}}
%
%terceira linha
%
 {\cal T}^{+} (\bar{\eta}, \eta; \nu; \sigma) &\equiv
\sum\limits_{l=1}^N \bar{\eta}_{[{(1-\sigma)\over 2}n +\nu]N + l}\;
\eta_{[{(1-\sigma)\over 2}n +\nu]N + l-1}\;\; . & (3.13c)\cr
}$$

We call \par 

$$\eqalignno{ {\cal E} (\bar{\eta}, \eta; \nu) &\equiv
\sum\limits_{\sigma=\pm 1} E(\sigma) {\cal E}(\bar{\eta}, \eta; \nu;
\sigma), & (3.14a) \cr
%
%segunda linha
%
 {\cal T}^{-} (\bar{\eta},
\eta; \nu) &\equiv \sum\limits_{\sigma\pm 1} t\, {\cal
T}^{-}(\bar{\eta}, \eta; \nu; \sigma), & (3.14b) \cr
%
%terceira linha
%
 {\cal T}^{+} (\bar{\eta},
\eta; \nu) &\equiv \sum\limits_{\sigma\pm 1} t \,{\cal
T}^{+}(\bar{\eta}, \eta; \nu; \sigma), & (3.14c) \cr
\noalign{\hbox{and}}
%
%quarta linha
%
 {\cal U}(\bar{\eta}, \eta; \nu)
&\equiv \sum\limits_{l=1}^N \bar{\eta}_{(n+\nu)N + l}\;\;
\eta_{(n+\nu)N + l}\;\; \bar{\eta}_{\nu N + l} \; \; \eta_{\nu N + l}.
& (3.14d) \cr   }$$ 

\noindent where $E(\sigma) \equiv E_0 - \sigma\lambda_B - \mu$. The
term ${\cal E} (\bar{\eta}, \eta; \nu)$ represents the kinetic
energy, ${\cal T}^{-} (\bar{\eta}, \eta; \nu)$ and $ {\cal T}^{+}
(\bar{\eta}, \eta; \nu)$ are the hopping terms and ${\cal U}
(\bar{\eta}, \eta; \nu)$ is the fermionic interaction term. \par 

Using the definitions (3.13) and (3.14), the Grassmann function ${\cal
K}^{\normal} (\bar{\eta}, \eta; \nu)$ is written as, \par 

$${\cal K}^{\normal} (\bar{\eta}, \eta; \nu)= {\cal E} (\bar{\eta},
\eta; \nu)+ {\cal T}^{-} (\bar{\eta}, \eta; \nu)+ {\cal T}^{+}
(\bar{\eta}, \eta; \nu) + {\cal U} (\bar{\eta}, \eta; \nu).
\eqno(3.15) $$ 

Since the terms in ${\cal E} (\bar{\eta}, \eta; \nu), {\cal T}^{-}
(\bar{\eta}, \eta; \nu), {\cal T}^{+} (\bar{\eta}, \eta; \nu)$ and
${\cal U} (\bar{\eta}, \eta; \nu)$ are products of $\bar{\eta}$ and
$\eta$ at the temperature index $\nu$ these expressions still have the
same form under the change of variables (3.6), except that they are
defined at $\nu+\nu_0$; that is, \par 

$$\eqalignno{ {\cal E} (\bar{\eta}, \eta; \nu) \hskip 0.3cm
&\rightarrow \hskip 0.3cm {\cal E} (\bar{\eta}, \eta; \nu+\nu_0) &
(3.16a) \cr
%
%segunda linha
%
 {\cal T}^{-} (\bar{\eta}, \eta; \nu)
\hskip 0.3cm &\rightarrow \hskip 0.3cm {\cal T}^{-} (\bar{\eta}, \eta;
\nu+\nu_0) & (3.16b) \cr
%
%terceira linha
%
 {\cal T}^{+} (\bar{\eta},
\eta; \nu) \hskip 0.3cm &\rightarrow \hskip 0.3cm {\cal T}^{+}
(\bar{\eta}, \eta; \nu+\nu_0) & (3.16c) \cr
%
%quarta linha
%
 {\cal U}
(\bar{\eta}, \eta; \nu) \hskip 0.3cm &\rightarrow \hskip 0.3cm 
{\cal U} (\bar{\eta}, \eta; \nu+\nu_0). & (3.16d) \cr }$$ 

Let us see the behavior of the expressions ${\cal E} (\bar{\eta},
\eta; \nu), {\cal T}^{-} (\bar{\eta}, \eta; \nu), {\cal T}^{+}
(\bar{\eta}, \eta; \nu)$ and \break ${\cal U} (\bar{\eta}, \eta; \nu)$
under the chiral transformations (3.8). We begin by considering ${\cal
E} (\bar{\eta}, \eta; \nu)$. From (3.13a) and (3.14a), we have that \par 

$$\eqalignno{ {\cal E} (\bar{\eta}, \eta; \nu) &=
\sum\limits_{\sigma=\pm 1} E(\sigma) \bar{\eta}_{[{(1-\sigma)\over 2}n
+\nu]N + 1}\; \eta_{[{(1-\sigma)\over 2}n +\nu]N + 1} + & \cr
%
%segunda linha
%
 &+ \sum\limits_{\sigma=\pm 1}\sum\limits_{l=2}^N
E(\sigma) \bar{\eta}_{[{(1-\sigma)\over 2}n +\nu]N + l}\;
\eta_{[{(1-\sigma)\over 2}n +\nu]N + l}\; . & (3.17) \cr }$$ 

The transformation (3.8) for a sector $\sigma\sigma$ is, \par 

$$\eqalignno{ 
\eta_{[{(1-\sigma)\over 2}n +\nu]N +1} = 
\eta_{[{(1-\sigma)\over 2}n +\nu]N +1}^\prime 
\hskip 0.5cm &{\rm and} \hskip 0.5cm
\bar{\eta}_{[{(1-\sigma)\over 2}n +\nu]N + 1} =
\bar{\eta}_{[{(1-\sigma)\over 2}n +\nu]N + 1}^\prime \;, & \cr & &
(3.18a) \cr
 \noalign{\hbox{for $l=1$ and $\nu= 0, 1, \cdots, n-1$,
and }}
%
%segunda linha
%
 \eta_{[{(1-\sigma)\over 2}n +\nu]N + l} =
\eta_{[{(1-\sigma)\over 2}n +\nu]N + N+2-l}^\prime
 \hskip 0.5cm
&{\rm and} \hskip 0.5cm
 \bar{\eta}_{[{(1-\sigma)\over 2}n +\nu]N +l} = 
\bar{\eta}_{[{(1-\sigma)\over 2}n +\nu]N + N+2-l}^\prime \;, &
\cr & & (3.18b) \cr }$$ 

\noindent for $l\not= 1$ and $\nu= 0, 1, \cdots, n-1$. \par 

Substituting (3.18) in eq.(3.17), and defining the index $k=N+2-l$ on
the second term in the r.h.s. of eq.(3.17), we obtain 

$$\eqalignno{ {\cal E} (\bar{\eta}, \eta; \nu) &=
\sum\limits_{\sigma=\pm 1} E(\sigma) \bar{\eta}_{[{(1-\sigma)\over 2}n
+\nu]N + 1}^\prime\; \eta_{[{(1-\sigma)\over 2}n +\nu]N + 1} ^\prime+
& \cr
%
%segunda linha
%
 &+ \sum\limits_{\sigma=\pm
1}\sum\limits_{k=2}^N E(\sigma) \bar{\eta}_{[{(1-\sigma)\over 2}n
+\nu]N + k}^\prime \; \eta_{[{(1-\sigma)\over 2}n +\nu]N + k}^\prime
\; . & \cr
%
%terceira linha
%
&= {\cal E} (\bar{\eta}^\prime,
\eta^\prime; \nu) & (3.19) \cr }$$ 

\noindent ${\cal E} (\bar{\eta}, \eta; \nu)$ is invariant under the
chiral transformation (3.18). \par 

Proceeding in a similar way, we also get that \par 

$$ {\cal U} (\bar{\eta}, \eta; \nu) = {\cal U} (\bar{\eta}^\prime,
\eta^\prime; \nu), \eqno(3.20) $$ 

\noindent where the relation between the generators $\{\eta,
\bar{\eta}\}$ and $\{\eta^\prime, \bar{\eta}^\prime\}$ is given by
(3.18). \par 

Let us now take under consideration one of the hopping terms, namely,
\par 

$$\eqalignno{ {\cal T}^{-} (\bar{\eta}, \eta; \nu) &=
\sum\limits_{\sigma=\pm 1} t \, \bar{\eta}_{[{(1-\sigma)\over 2}n
+\nu]N + 1}\; \eta_{[{(1-\sigma)\over 2}n +\nu]N + 2} + & \cr
%
%segunda linha
%
 &+ \sum\limits_{\sigma=\pm 1}\sum\limits_{l=2}^{N-1}
t \, \bar{\eta}_{[{(1-\sigma)\over 2}n +\nu]N + l}\;
\eta_{[{(1-\sigma)\over 2}n +\nu]N + l+1}\ + & \cr
%
%terceira linha
%
&+ \sum\limits_{\sigma=\pm 1} t \,
 \bar{\eta}_{[{(1-\sigma)\over 2}n +\nu]N + N}\; 
\eta_{[{(1-\sigma)\over 2}n +\nu]N + N+1} \;\;, & (3.21)
\cr }$$ 

\noindent where we have already taken into account the periodic space
boundary condition. \par 

Applying the change of variables (3.18) in eq.(3.21) and defining the
index $k=N+2-l$, we have

$$\eqalignno{ {\cal T}^{-} (\bar{\eta}, \eta; \nu) &=
\sum\limits_{k=1}^N \sum\limits_{\sigma=\pm 1}
\bar{\eta}_{[{(1-\sigma)\over 2}n +\nu]N + k}\;
\eta_{[{(1-\sigma)\over 2}n +\nu]N + k-1} & \cr
%
%segunda linha
%
 &= {\cal T}^{+} (\bar{\eta}, \eta; \nu) & (3.22) \cr }$$ 

In a similar way, it may be shown that, under (3.18), ${\cal T}^{+}
(\bar{\eta}, \eta; \nu)$ transforms into \break
 ${\cal T}^{-} (\bar{\eta}, \eta; \nu)$. \par 

We point out that the results of this section are valid for any
positive integer values of $n$ and $N$. \par 

\vskip 1cm 

\noindent {\bf 4. Calculation of the Coefficients of the Grand
Canonical Partition Function for the Hubbard Model in the High
Temperature Limit} 

\bigskip 

Up to now, all that we have said about ${\rm Tr} [{\bf K}^n]$ in
sections 2 and 3, is valid for any value of $n$ and $N$. \par 

The calculation of ${\rm Tr} [{\bf K}^n]$, given by eq.(2.15), for
arbitrary $n$ can formally be done using the result (2.20). However,
due to the fact that the number of terms that contribute to ${\rm Tr}
[{\bf K}^n]$ increases rapidly with $n$, we are able to calculate the
exact coefficients up to order $\beta^3$ of the grand canonical
partition function of unidimensional Hubbard model in the high
temperature limit. \par 

When we say that our result is exact up to order $\beta^3$, we mean
that it is valid for any value of the constant parameters $E_0, t, U$
and $\mu$, that characterize the model, and, for any value of the
constant external magnetic field $B$. \par 

\vskip 1cm 

\noindent {\bf 4.1. The Exact $\beta^2$ Coefficient in d=(1+1)
Hubbard Model} 

\bigskip 

From eq.(2.15), the expression of ${\rm Tr} [{\bf K}^2]$ in terms of
the Grassmann generators is \par 

$$ {\rm Tr}[{\bf K}^2] = \int \prod_{I=1}^{4N} \, d\eta_I
d\bar{\eta}_I \;\; e^{\sum\limits_{I,J= 1}^{4N} \bar{\eta}_I\; A_{I J}
\; \eta_J}
 \times {\cal K}^{\normal} (\bar{\eta}, \eta; \nu=0)\;
{\cal K}^{\normal} (\bar{\eta}, \eta; \nu=1) \;, \eqno (4.1.1) $$ 

\noindent where the matrix {\bf A} is given by (2.13a), and for the
unidimensional periodic Hubbard model, ${\cal K}^{\normal}$ is given
by eq.(2.1.3) or eq.(3.15). \par 

For $n=2$, the matrices ${\bf A}^{\uparrow\uparrow}$ and ${\bf
A}^{\downarrow\downarrow}$ (eq.(2.14a)) are: 

$$ {\bf A}^{\uparrow\uparrow} = {\bf A}^{\downarrow\downarrow} =
\pmatrix{ \lbarra_{N\times N} & - \lbarra_{N\times N} \!\!\cr & & \!\!
\cr \lbarra_{N\times N} & \lbarra_{N\times N} \!\! \cr}, \eqno(4.1.2)
$$ 

\noindent where $\lbarra_{N\times N}$ is the identity matrix of
dimension $N\times N$. For the case $n=2$, the eigenvalues of the
matrices ${\bf A}^{\uparrow\uparrow}$ and $ {\bf
A}^{\downarrow\downarrow}$ are: \par 

$$ \lambda_1 = \lambda_2^* = 1- e^{i\pi\over 2}. \eqno( 4.1.2a)$$ 

\noindent The eigenvalues $\lambda_1$ and $\lambda_2$ are $N$-fold
degenerated. \par 

>From eq.(2.18a), {\bf P} is the matrix that, through a similarity
transformation, diagonalizes the matrices ${\bf A}^{\uparrow\uparrow}$
and $ {\bf A}^{\downarrow\downarrow}$. For $n=2$, the matrix {\bf P}
is, \par 

$$ {\bf P} = \pmatrix{ \lbarra{N\times N} & \lbarra_{N\times N}
\!\!\cr & & \!\! \cr i\, \lbarra_{N\times N} & -i\, \lbarra_{N\times
N} \!\! \cr}, \eqno(4.1.2b) $$ 

\noindent and its inverse is: \par 

$$ {\bf P}^{-1} = {1\over 2} \pmatrix{ \lbarra_{N\times N} & - i
\,\lbarra_{N\times N} \!\!\cr & & \!\! \cr
 \lbarra_{N\times N} & i \,
\lbarra_{N\times N} \!\! \cr}. \eqno(4.1.2c) $$ 

Since {\bf P} and ${\bf P}^{-1}$ are block--matrices, their entries
can be written as $P_{\nu N +l, \nu^\prime N +k} = \delta_{lk} p_{\nu
\nu^\prime}$ and $P_{\nu N +l, \nu^\prime N +k}^{-1} = \delta_{lk}
q_{\nu \nu^\prime}$. $p_{\nu \nu^\prime}$ and $q_{\nu \nu^\prime}$ are
the non--null elements and for $n=2$ they are equal to \par 

$$p_{\nu \nu^\prime} = \pmatrix{ 1 & 1\cr i & - i\cr} \hskip 0.7cm
{\rm and} \hskip 0.7cm
 q_{\nu \nu^\prime}= {1\over 2} \pmatrix { 1 &
-i \cr 1 & i\cr }. \eqno(4.1.2d) $$ 

Therefore, \par 

$$ \eqalignno{ {\bf D} &\equiv {\bf P}^{-1} {\bf A}^{\sigma\sigma}
{\bf P} & \cr & & \cr
%
%segunda linha
%
 &= \pmatrix{ (1- e^{i\pi\over
2})\; \lbarra_{N\times N} & \0barra_{N\times N} \cr & & \cr
\0barra_{N\times N} & (1- e^{-i\pi\over 2})\; \lbarra_{N\times N}\cr
}, \hskip 0.5cm {\rm for} \;\;\; \sigma=\uparrow, \downarrow. &
(4.1.2e) \cr }$$ 

For $n=2$, the change of variables (2.19): $\eta= {\bf P}\eta^\prime$
and $\bar{\eta}= \bar{\eta}^\prime {\bf P}^{-1}$, becomes, \par 

$$\eqalignno{ \eta_{\nu N +l} & = \sum\limits_{\nu^\prime = 0}^1 \;
p_{\nu \nu^\prime} \; \eta_{\nu^\prime N +l}^\prime & (4.1.3a) \cr
\noalign{\hbox{and}}
%
%segunda linha
%
 \bar{\eta}_{\nu N +l} & =
\sum\limits_{\nu^\prime = 0}^1 \; \bar{\eta}_{\nu^\prime N +l}^\prime
\; q_{\nu^\prime \nu} \;, & (4.1.3b) \cr }$$ 

\noindent where $\nu=0,1$ and $l=1, 2, \cdots, N$. \par 

We should note that due to the block structure of the matrices {\bf P}
and ${\bf P}^{-1}$, the space index $l$ in the generators $\eta$ and
$\bar{\eta}$ are not mixed up by the transformation (4.1.3). \par 

From eq.(3.15), we write function ${\cal K}^{\normal}
(\bar{\eta},\eta;\nu)$ as \par 

$$ {\cal K}^{\normal} (\bar{\eta},\eta;\nu) = {\cal E}
(\bar{\eta},\eta;\nu) + {\cal T}^{-} (\bar{\eta},\eta;\nu) + 
{\cal T}^{+} (\bar{\eta},\eta;\nu) + {\cal U} (\bar{\eta},\eta;\nu).
\eqno(4.1.4) $$ 

In order to simplify the notation, we define \par 

$$\eqalignno{ < {\cal O}_1 (\nu_1) \cdots {\cal O}_m (\nu_m) > &
\equiv \int \prod_{I=1}^{2nN} \, d\eta_I d\bar{\eta}_I \;\;
e^{\sum\limits_{I,J= 1}^{2nN} \bar{\eta}_I\; A_{I J} \; \eta_J} \;\;
\times & \cr
 & \hskip 1cm \times \;\; {\cal O}_1
(\bar{\eta},\eta;\nu_1) \cdots {\cal O}_m (\bar{\eta},\eta;\nu_m) &
(4.1.5a) \cr
 \noalign{\hbox{and}}
%
%segunda linha
%
 < {\cal O}_1
(\nu_1) \cdots {\cal O}_m (\nu_m) >_\sigma &\equiv \int
\prod_{I=(1-\sigma)nN+1}^{(3-\sigma)nN}\, d\eta_I d\bar{\eta}_I \;\;
e^{\sum\limits_{I,J= 1}^{2nN} \bar{\eta}_{(1-\sigma)nN+I} \; A_{I J}
\; \eta_{(1-\sigma)nN+J}} \;\; \times & \cr
 & \hskip 1cm \times
{\cal O}_1 (\bar{\eta},\eta;\nu_1) \cdots {\cal O}_m
(\bar{\eta},\eta;\nu_m) & (4.1.5b) \cr }$$ 

\noindent where ${\cal O}_j (\bar{\eta},\eta;\nu_j)$ are Grassmann
functions. \par 

When we substitute eq.(4.1.4) for $\nu=0$ and $\nu=1$ in eq.(4.1.1),
we get sixteen terms. However, using the results (3.19), (3.20) and
(3.22), and the fact that the terms: \break $<{\cal E}(0) {\cal
T}^-(1)>, <{\cal E}(0) {\cal T}^+(1)>, <{\cal T}^-(0) {\cal T}^-(1)>,
<{\cal T}^-(0) {\cal U}(1)>, <{\cal T}^+(0) {\cal T}^+(1)>$, and
$<{\cal T}^+(0) {\cal U}(1)>$ are null, we have that
 ${\rm Tr}[{\bf K}^2]$ reduces to the sum of four distinct 
terms: \par 

$$ {\rm Tr}[{\bf K}^2] = <{\cal E}(0) {\cal E}(1)> + 2 <{\cal T}^-(0)
{\cal T}^+(1)> + 2<{\cal E}(0) {\cal U}(1)> + <{\cal U}(0) {\cal
U}(1)>. \eqno(4.1.6) $$ Before calculating these terms, lets us
consider one of the null integrals and show the reasoning for its
vanishing. For example, the term \par 

$$ <{\cal E}(0) {\cal T}^-(1)> = 2 t (E_0 - \mu) [ <{\cal E}(\uparrow,
0) {\cal T}^-(\uparrow, 1)> + <{\cal E}(\uparrow, 0) {\cal
T}^-(\downarrow, 1)> ]. \eqno(4.1.7) $$ 

In expression (4.1.7) we identify the following types of terms: \par 

i) Terms coming from $<{\cal E}(\uparrow, 0) {\cal T}^-(\uparrow,
1)>$: \par 

$$\eqalignno{ {\cal I}_1 &\equiv \int \prod_{I=1}^{2N} \, d\eta_I
d\bar{\eta}_I\;\; e^{\sum\limits_{I,J= 1}^{2N} \bar{\eta}_I\; A_{I J}
\; \eta_J} \;\; \bar{\eta}_{l_1} \eta_{l_1} \bar{\eta}_{N+ l_2}
\eta_{N+l_2 +1} \; \; \times & \cr
%
%segunda linha
%
 & \hskip 3cm
\times \int \prod_{I=2N+1}^{4N} \, d\eta_I d\bar{\eta}_I \;\;
e^{\sum\limits_{I,J= 2N+1}^{4N} \bar{\eta}_I\; A_{I J} \; \eta_J}, &
(4.1.7a) \cr }$$ 

\noindent where $l_1$ and $l_2$ are fixed and represent the space site
index. \par 

Making the change of variables (4.1.3) to diagonalize ${\bf A}^{\sigma
\sigma}$, $\sigma= \uparrow, \downarrow$, we get \par 

$$ \eqalignno{ {\cal I}_1 & = \det({\bf A}^{\downarrow\downarrow}) \;
\sum\limits_{ \nu_1, \nu_1^\prime =0 \atop{ \nu_2, \nu_2^\prime}}^1 \;
q_{{\nu_1} 0} p_{0 {\nu_1^\prime}} q_{{\nu_2} 1} p_{1 {\nu_2^\prime}}
\; \times & \cr
%
%segunda linha
%
 & \times \; \int \prod_{I=1}^{2N}
\, d\eta_I^\prime d\bar{\eta}_I^\prime \;\; e^{\sum\limits_{I,J=
1}^{2N} \bar{\eta}_I^\prime\; D_{I J} \; \eta_J^\prime} \;\;
\bar{\eta}_{\nu_1 N +l_1}^\prime \eta_{\nu_1^\prime N + l_1}^\prime
\bar{\eta}_{\nu_2 N+ l_2}^\prime \eta_{\nu_2^\prime N+l_2 +1}. &
(4.1.7b) \cr }$$ 

The presence of $\bar{\eta}_i$ and $\eta_j$ in the integrand of
eq.(4.17b) implies that the integral is equal to the determinant of a
matrix obtained from matrix {\bf D} by cutting the line $i$ and the
column $j$. As long as {\bf D} is diagonal, the only non--vanishing
integrals are those for which $i=j.$ In other words, if the $i$-th
line of {\bf D} is cut, so must be its $i$-th column. \par 

In eq.(4.1.7b) the space index $(l_2 + 1)$ {\bf never} equals any
other space index. Therefore, the minor obtained from matrix {\bf D}
by cutting the lines: $(\nu_1 + l_1)$ and $(\nu_2 + l_2)$, and the
columns: $(\nu_1^\prime + l_1)$ and $(\nu_2^\prime + l_2 +1)$, will
always have, at least, one line or row of zeros. Therefore, its
determinant vanishes. \par 

Then, \par 

$$ {\cal I}_1 =0 \Rightarrow <{\cal E}(\uparrow, 0) {\cal
T}^-(\uparrow, 1)> = 0. \eqno(4.1.8a)$$ 

\vskip 0.3cm 

ii) Terms coming from $<{\cal E}(\uparrow, 0) {\cal T}^-(\downarrow,
1)>$: \par 

Using a similar reasoning as for the terms in $<{\cal
E}(\uparrow, 0) {\cal T}^-(\uparrow, 1)>$ , we get that \par 

$$ < {\cal T}^- (\downarrow, 1)>_1 = 0. \eqno(4.1.8b)$$ 

Therefore, using the results (4.1.8a--b), we finally obtain, \par 

$$ <{\cal E}(0) {\cal T}^-(1)> = 0. \eqno(4.1.9) $$ 

Proceeding in an analogous way, it is straightforward to show that the
terms \break $<{\cal E}(0) {\cal T}^+(1)>, <{\cal T}^-(0) {\cal
T}^-(1)>, <{\cal T}^-(0) {\cal U}(1)>, <{\cal T}^+(0) {\cal T}^+(1)>$
and $ <{\cal T}^+(0) {\cal U}(1)>$ also vanish. \par 

The results (4.1.8a--b) are a direct consequence to the fact that the
change of variables (4.1.3) does not mix up space indices. \par 

Taking into account the previous results and calculating the
coefficients of the terms of eq.(4.1.6), ${\rm Tr}[{\bf K}^2]$ may be
written as: \par 

$$\eqalignno{ {\rm Tr}[{\bf K}^2] &= 2( (E_0 - \mu)^2 +\lambda_B^2)
<{\cal E}(\uparrow,0) {\cal E}(\uparrow,1)> + 2 ( (E_0 - \mu)^2
-\lambda_B^2) <{\cal E}(\downarrow,0)> <{\cal E}(\uparrow,0)> + & \cr
%
% segunda linha
%
 &+ 4 t^2 <{\cal T}^-(\uparrow,0) {\cal
T}^+(\uparrow,1)>+ 4 (E_0 - \mu) <{\cal E}(\uparrow,0) {\cal U}(1)> +
<{\cal U}(0) {\cal U}(1)>. & \cr & & (4.1.10) \cr }$$ 

The terms on the r.h.s. of eq.(4.1.10) are sums of multivariable
Grassmann integrals that reduce to the following types: \par 

$$\eqalignno{ {\cal H}_1 (l) &\equiv \int \prod_{I=1}^{2N} \, d\eta_I
d\bar{\eta}_I \;\; e^{\sum\limits_{I,J= 1}^{2N} \bar{\eta}_I\; A_{I
J}^{\uparrow \uparrow} \; \eta_J} \;\; \bar{\eta}_l \;\eta_l , &
(4.1.11a) \cr
%
%segunda linha
%
 {\cal H}_2 (l,k) &\equiv \int
\prod_{I=1}^{2N} \, d\eta_I d\bar{\eta}_I \;\; 
e^{\sum\limits_{I,J=1}^{2N}
 \bar{\eta}_I\; A_{I J}^{\uparrow \uparrow} \; \eta_J} \;\;
\bar{\eta}_l\; \eta_l \;\;\bar{\eta}_{N+k} \;\eta_{N+k}, &
(4.1.11b)\cr
 \noalign{\hbox{and}}
%
%terceira linha
%
 {\cal H}_3
(l,k) &\equiv \int \prod_{I=1}^{2N} \, d\eta_I d\bar{\eta}_I \;\;
e^{\sum\limits_{I,J= 1}^{2N} \bar{\eta}_I\; A_{I J}^{\uparrow
\uparrow} \; \eta_J} \;\; \bar{\eta}_l \;\eta_{l+1}
\;\;\bar{\eta}_{N+k}\; \eta_{N+k-1}, & (4.1.11c)\cr }$$ 

\noindent in the sector $\sigma\sigma= \uparrow\uparrow$, and
equivalent integrals in the sector $\downarrow\downarrow$. \par 

Let us calculate in detail these integrals. Those integrals that
contribute to order $\beta^3$ of the expansion of the grand canonical
partition function in the high temperature limit are handled in 
a similar way. In the next section, we will restrict ourselves 
to give the results of such integrals. \par 

Under the change of variables (4.1.3), the integral ${\cal H}_1(l)$
becomes: \par 

$$ {\cal H}_1(l) = \sum\limits_{\nu, \nu^\prime=0}^1 q_{\nu 0} p_{0
\nu^\prime} \int \prod_{I=1}^{2N} \, d\eta_I^\prime
d\bar{\eta}_I^\prime \;\; e^{\sum\limits_{I,J= 1}^{2N}
\bar{\eta}_I^\prime\;D_{I J}\;\eta_J^\prime} \;\; \bar{\eta}_{\nu
N+l}^\prime \;\eta_{\nu^\prime N+l}^\prime. \eqno(4.1.12) $$ 

We only get non--null results for eq.(4.1.12) when the $i$-th line and
the $i$-th column of the matrix {\bf D} are cut simultaneously. This
only happens when $\nu=\nu^\prime$. \par 

Therefore, \par 

$$ {\cal H}_1 (l) = \int \prod_{I=1}^{2N} \, d\eta_I^\prime
d\bar{\eta}_I^\prime \;\; e^{\sum\limits_{I,J= 1}^{2N}
\bar{\eta}_I^\prime\;D_{I J}\;\eta_J^\prime} \;\; ( q_{00} p_{00}
\;\bar{\eta}_l^\prime \eta_l^\prime + q_{10} p_{01}
\;\bar{\eta}_{N+l}^\prime \eta_{N+l}^\prime). \eqno(4.1.13) $$ 

\noindent The first term in the integrand corresponds to cut the
$l$-th line and $l$-th column of matrix {\bf D}. Since it is a
diagonal block--matrix, the result of the integral is independent of
$l$ and it is equal to: $\lambda_1^{N-1} \lambda_2^{N}$. The second
term in the integrand corresponds to cutting its  $(N+l)$-th line and
$(N+l)$-th column. By the same reason as before, the result of this
integral is independent of the value of $l$, and it is equal to
$\lambda_1^{N} \lambda_2^{N-1}$. \par 

Substituting the values of $q_{ij}$ and $p_{ij}$ (eq.(4.1.2d)) in
eq.(4.1.13), we finally get \par 

$$ \eqalignno{ {\cal H}_1 (l) &\equiv \int \prod_{I=1}^{2N} \, d\eta_I
d\bar{\eta}_I \;\; e^{\sum\limits_{I,J= 1}^{2N} \bar{\eta}_I\; A_{I
J}^{\uparrow\uparrow} \; \eta_J} \;\; \bar{\eta}_l \;\eta_l , & \cr
%
%segunda linha
%
 & = 2^{N-1}, & (4.1.14) \cr }$$ 

\noindent for any value of $l$. \par 

Making the change of variables (4.1.3) in the integral ${\cal
H}_2(l,k)$, it becomes: \par 

$$\eqalignno{ {\cal H}_2 (l,k) & = \sum_{\nu_1, \nu_1^\prime =0
\atop{\nu_2, \nu_2^\prime =0 }}^1 \; q_{\nu_1 0} p_{0 \nu_1^\prime} \,
q_{\nu_2 1} p_{1 \nu_2^\prime} \;\; \times & \cr
%
%segunda linha
%
& \hskip -0.5cm \times \int \prod_{I=1}^{2N} \, d\eta_I^\prime
d\bar{\eta}_I^\prime \;\; e^{\sum\limits_{I,J= 1}^{2N}
\bar{\eta}_I^\prime\;D_{I J}\;\eta_J^\prime} \;\; \bar{\eta}_{\nu_1
N+l}^\prime\; \eta_{\nu_1^\prime N+l}^\prime \;\;\bar{\eta}_{\nu_2
N+k}^\prime \;\eta_{\nu_2^\prime N+k}^\prime. & (4.1.15a) \cr }$$ 

\noindent Once the $i$-th line and the $i$-th column need to be cut
simultaneously if integral (4.1.15a) is to be non-vanishing, we have
two situations to consider: \par 

$i) l \not= k.$ \par 

In this case, the non--zero terms of eq.(4.1.15a) are: \par 

$$\eqalignno{ {\cal H}_2 (l,k) & = \sum\limits_{\nu_1, \nu_2=0 }
q_{\nu_1 0} p_{0 \nu_1} \, q_{\nu_2 1} p_{1 \nu_2 } \; \; \times & \cr
%
%segunda linha
%
& \hskip -1cm
 \times \int \prod_{I=1}^{2N} \, d\eta_I^\prime
d\bar{\eta}_I^\prime \;\; e^{\sum\limits_{I,J= 1}^{2N}
\bar{\eta}_I^\prime\;D_{I J}\;\eta_J^\prime} \;\; \bar{\eta}_{\nu_1
N+l}^\prime\; \eta_{\nu_1 N+l}^\prime \;\;\bar{\eta}_{\nu_2
N+k}^\prime \;\eta_{\nu_2 N+k}^\prime. & (4.1.15b) \cr } $$ 

\noindent Since the result of the integrals in eq.(4.1.15b) i{
independent of the space indices $l$ and $k$ ($l\not= k$), only the
total number of cuts within each sector $\nu$ is relevant to their
evaluation. For example, for $\nu_1=0$ and $\nu_2=0$, there are two
cuts in the sector $\nu=0$ of matrix {\bf D}. Thus the result of this
integral is $\lambda_1^{N-2} \lambda_2^N$. For $\nu_1=0$ and
$\nu_2=1$, there is one cut in the sector $\nu=0$ and another one in
the sector $\nu=1$ of matrix {\bf D}; hence, the integral is equal to
$\lambda_1^{N-1} \lambda_2^{N-1}$, the same result as for the case
$\nu_1=1$ and $\nu_2=0$. \par 

Using the previous reasoning and the values of $q_{ij}$ and $p_{ij}$
given by eq.(4.1.2d), we get, \par 

$$ {\cal H}_2 (l,k) = 2^{N-2}, \hskip 0.5cm {\rm for} \hskip 0.5cm l,
k= 1, 2, \cdots, N \hskip 0.5cm {\rm and} \hskip 0.5cm l\not= k.
\eqno(4.1.15c) $$ 

\vskip 0.3cm 

$ ii) l=k.$ \par 

In this case, we must take into account that $\eta_i^2 =
\bar{\eta}_i^2=0$. For $l=k$ we have to compute all the possible
permutations of the product $\bar{\eta}_l^\prime \eta_l^\prime \,
\bar{\eta}_{N+l}^\prime \eta_{N+l}^\prime$. The only non--null terms
in eq.(4.1.15a) are: \par

$$\vbox{ \halign{ \hskip 0.3cm # \hskip 0.3cm & # \hskip 0.3cm & #
\hskip 0.3cm &# \cr ${\bf q}_{\nu_1 0}$ & ${\bf p}_{0 \nu_1}$ & ${\bf
q}_{\nu_2 1}$ & ${\bf p}_{1 \nu_2}$ \cr
\noalign{\smallskip \hrule \smallskip} 

$\bar{\eta}_l$ & $\eta_l$ & $\bar{\eta}_{N+l}$ & $\eta_{N+l}$ \cr
$\bar{\eta}_l $ & $\eta_{N+l}$ & $\bar{\eta}_{N+l}$ & $\eta_l$ \cr
$\bar{\eta}_{N+l}$ & $\eta_{N+l}$ & $\bar{\eta}_l$ & $\eta_l$ \cr
$\bar{\eta}_{N+l}$ & $\eta_l $ & $\bar{\eta}_l$ & $\eta_{N+l}$ \cr } }
\vbox {\+ \phantom{...} \cr \+ $ \;\;\Rightarrow$ \hskip 0.5cm $
q_{00}\,p_{00} \,q_{11}\, p_{11} $ \cr
 \+ $\;\;\Rightarrow$ \hskip
0.5cm $ q_{00}\,p_{01} \, q_{11}\, p_{10}$ \cr
 \+ $\;\;\Rightarrow$
\hskip 0.5cm $ q_{10}\,p_{01} \, q_{01}\, |_{10}$ \cr
 \+ $\;\;\Rightarrow$ \hskip 0.5cm $ q_{10}\,p_{00} \,q_{01}\, p_{11}$ \cr
} \eqno(4.1.16)$$ 

To write the multivariable Grassmann integrals as co--factors of
matrix {\bf D}, we have to rearrange the product to the pattern $
\bar{\eta}_l^\prime \eta_l^\prime \, \bar{\eta}_{N+l}^\prime
\eta_{N+l}^\prime$. Because of the permutations, the second and
fourth term in (4.1.16) get a minus sign. \par 

By the same reason as explained in the calculation of integral ${\cal
H}_1 (l)$, the result of the integrals is independent of $l$, and it
is equal to \par 

$$ {\cal H}_2 (l,l) = 2^{N-1}, \hskip 0.5cm {\rm for} \hskip 0.5cm
l=1, 2, \cdots, N. \eqno(4.1.17) $$ 

In summary, we have that \par 

$$\eqalignno{ {\cal H}_2 (l,k) &\equiv \int \prod_{I=1}^{2N} \,
d\eta_I d\bar{\eta}_I \;\; e^{\sum\limits_{I,J= 1}^{2N} \bar{\eta}_I\;
A_{I J}^{\uparrow\uparrow} \; \eta_J} \;\; \bar{\eta}_l\; \eta_l
\;\;\bar{\eta}_{N+k} \;\eta_{N+k}, & \cr
%
%outras linhas
%
&= \cases{ 2^{N-2}, & if $l\not= k$ \cr 2^{N-1}, & if $l=k$ \cr 0, &
otherwise .\cr } & (4.1.18) \cr }$$ 

Finally, we calculate the integral ${\cal H}_3(l,k)$. Under the change
of variables (4.1.3), it becomes: \par 

$$\eqalignno{ {\cal H}_3 (l,k) &\equiv - \sum_{ \nu_1, \nu_1^\prime=0
\atop { \nu_2, \nu_2^\prime=0 }}^1 \; q_{\nu_1 0} \, p_{0
\nu_1^\prime} \, q_{\nu_2 1} \, p_{1 \nu_2^\prime} \;\; \times & \cr
%
%segunda linha
%
 & \times \int \prod_{I=1}^{2N} \, d\eta_I^\prime
d\bar{\eta}_I^\prime \;\; e^{\sum\limits_{I,J= 1}^{2N}
\bar{\eta}_I^\prime\;D_{I J}\;\eta_J^\prime}
 \;\; \bar{\eta}_{\nu_1
N+l}^\prime \;\eta_{\nu_1^\prime N+k-1}^\prime \;\;\bar{\eta}_{\nu_2
N+k}^\prime\; \eta_{\nu_2^\prime N+l+1}^\prime. & \cr & & (4.1.19) \cr
}$$ 

The only non--null integrals in ${\cal H}_3 (l,k)$ are the ones where
$l=k-1$. The terms on the r.h.s. of (4.1.19) that contribute to it are
those where $\nu_1 = \nu_1^\prime$ and $\nu_2 = \nu_2^\prime$. \par 

Using the values of $q_{ij}$ and $p_{ij}$ given by eq.(4.1.2d), we get
\par 

$$ {\cal H}_3(l, l+1) = 2^{N-2}, \hskip 0.5cm l=1, 2, \cdots, N.
\eqno(4.1.20) $$ 

Therefore, we have that, \par 

$$\eqalignno{ {\cal H}_3 (l,k) &\equiv \int \prod_{I=1}^{2N} \,
d\eta_I d\bar{\eta}_I \;\; e^{\sum\limits_{I,J= 1}^{2N}
\bar{\eta}_I\;A_{I J}^{\uparrow \uparrow} \;\eta_J} \;\; \bar{\eta}_l
\;\eta_{l+1} \;\;\bar{\eta}_{N+k}\; \eta_{N+k-1}. & \cr
%
%outras linhas
%
 & & \cr
 & =\cases { 2^{N-2},& if $k=l+1$ \cr 
	    0,& if $k\not= l+1$ \cr } & (4.1.20a) \cr
 } $$ 

From the results (4.1.14), (4.1.18) and (4.1.20a) we evaluate the
terms $ <{\cal E}(\uparrow, 0) {\cal E}(\uparrow,1)>, \break
 <{\cal E}(\sigma, 0)>_\sigma,
 <{\cal T}^-(\uparrow, 0) {\cal T}^+(\uparrow,1)>,
<{\cal E}(\uparrow, 0) {\cal U}(1)>$ and $<{\cal U}(0) {\cal U}(1)>$.
Substituting the results in (4.1.10), we finally get: \par 

$$ {\rm Tr} [{\bf K}^2] = 2^{2N} \big[ N^2 \big( (E_0 - \mu)^2 + 
{1\over 2}(E_0 - \mu) U + {U^2 \over 16} \big) +
 {N\over 2} \big( (E_0 - \mu)^2 + (E_0 - \mu) U + 
\lambda_B^2 + 2 t^2 + {3 \over 8} U^2 \big) \big],
\eqno(4.1.21) $$ 

\noindent where $\lambda_B = -{1\over 2} g\mu_B B$. \par 

The result (4.1.21) is valid for any value of $N$ space sites, and any
set of values: \break $(E_0, t, U, \mu)$ that characterize the
fermionic system and its interaction with the external magnetic field
$B$. \par 

\vskip 1cm 

\noindent {\bf 4.2. The Exact $\beta^3$ Coefficient in d=(1+1)
Hubbard Model} 

\bigskip 

In section 4.1 we used the simpler case $\beta^2$ to exemplify, in
great detail, how to use the results of Appendix A to calculate the
coefficients of the expansion (2.4). In this section, we present the
results of ${\rm Tr}[{\bf K}^3]$ for any number $N$ of space sites.
\par 

The expression of (3.12) for $n=3$ is \par 

$$\eqalignno{ {\rm Tr}[{\bf K}^3] &= \int \prod_{I=1}^{6N} \, d\eta_I
d\bar{\eta}_I \;\; e^{\sum\limits_{I,J= 1}^{6N} \bar{\eta}_I\; A_{I J}
\; \eta_J} \times & \cr
%
%segunda linha
%
 & \hskip -1cm \times {\cal
K}^{\normal} (\bar{\eta}, \eta; \nu=0)\; {\cal K}^{\normal}
(\bar{\eta}, \eta; \nu=1) \; {\cal K}^{\normal} (\bar{\eta}, \eta;
\nu=2), & (4.2.1a) \cr }$$ 

\noindent where $A_{IJ}$ aze(the entries of the block--matrix {\bf A}
(eq.(2.13a). For $n=3$, we have that \par 

$$ {\bf A}^{\uparrow \uparrow} = {\bf A}^{\downarrow \downarrow} =
\pmatrix { \lbarra_{N\times N} & - \lbarra_{N\times N} &
\0barra_{N\times N} \cr
 & & & \cr
%
%segunda linha da matriz
%
\0barra_{N\times N} & \lbarra_{N\times N}& - \lbarra{N\times N} \cr
%
%terceira linha da matriz
%
%
 & & & \cr
%
%ultima linha da matriz
%
\lbarra_{N\times N} & \0barra_{N\times N} & \lbarra_{N\times N} \cr}.
\eqno(4.2.1b) $$ 

The expression of the Grassmann function ${\cal K}^{\normal}
(\bar{\eta}, \eta; \nu)$ for the case of the periodic unidimensional
Hubbard model is given by eq.(2.1.3). \par 

The r.h.s. of eq.(4.2.1a) is expanded into a sum of $64$ terms, each
of which contains a Grassmann multivariable integral. Their evaluation
will be attained by the same technique used in section 4.1 for the
case $n=2$. \par

For $n=3$, the matrices $A^{\sigma\sigma}$, $\sigma= \uparrow,
\downarrow$, are diagonalized by the matrix {\bf P} (eq.(2.18a)),
\par 

$$ {\bf P} = \pmatrix{ \lbarra^{N\times N} & (1 + e^{2\pi i\over
3})\lbarra_{N\times N} & (1 + e^{-2\pi i\over 3})\lbarra_{N\times N}
\cr & & \cr
%
%segunda linha da matriz
%
 - \lbarra_{N\times N} &
\lbarra_{N\times N} & \lbarra_{N\times N} \cr & & \cr
%
%terceira linha da matriz
%
 \lbarra_{N\times N} & (1 + e^{-2\pi i\over
3})\lbarra_{N\times N} & (1 + e^{2\pi i\over 3})\lbarra_{N\times N}
\cr }, \eqno(4.2.2a) $$ 

\noindent the inverse of which is \par 

$$ {\bf P}^{-1} = {1\over 3} \pmatrix{ \lbarra_{N\times N} &
-\lbarra_{N\times N} & \lbarra_{N\times N}\cr & & \cr
%
%segunda linha da matriz
%
 (1 + e^{-2\pi i\over 3})\lbarra_{N\times N}&
\lbarra_{N\times N}& (1 + e^{2\pi i\over 3})\lbarra_{N\times N} \cr &
& \cr
%
%terceira linha da matriz
%
 (1 + e^{2\pi i\over
3})\lbarra_{N\times N} & \lbarra_{N\times N} & (1 + e^{-2\pi i\over
3})\lbarra{N\times N} \cr }. \eqno (4.2.2b) $$ 

For $n=3$, the diagonalized matrix {\bf D} (eq.(2.18)) is \par 

$$ {\bf D} = \pmatrix{ 2 \lbarra_{N\times N} & \0barra_{N\times N} &
\0barra_{N\times N} \cr & & \cr
%
%segunda linha da matriz
%
\0barra_{N\times N} & (1 + e^{2\pi i\over 3})\lbarra_{N\times N}&
\0barra_{N\times N} \cr & & \cr
%
%terceira linha da matriz
%
\0barra_{N\times N} & \0barra_{N\times N} & (1 + e^{-2\pi i\over
3})\lbarra_{N\times N} \cr }. \eqno(4.2.3) $$ 

\noindent The eigenvalues of the matrix ${\bf A}^{\sigma\sigma}$,
$\sigma= \uparrow, \downarrow$, are: $ 2, (1 + e^{2\pi i\over 3})$ and
$(1 + e^{-2\pi i\over 3})$. Each eigenvalue is $N$--fold degenerated.
\par 

The change of variables (2.19), for the case $n=3$ becomes, \par 

$$\eqalignno{ \eta_{\nu N +l} & = \sum\limits_{\nu^\prime = 0}^2 \;
p_{\nu \nu^\prime} \; \eta_{\nu^\prime N +l}^\prime & (4.2.4a) \cr
\noalign{\hbox{and}}
%
%segunda linha
%
 \bar{\eta}_{\nu N +l} & =
\sum\limits_{\nu^\prime = 0}^2 \; \bar{\eta}_{\nu^\prime N +l}^\prime
\; q_{\nu^\prime \nu} \;, & (4.2.4b) \cr }$$ 

\noindent with $l=1, 2, \cdots,N$, and \par 

$$ \eqalignno{ p_{\nu^\prime \nu} & = \pmatrix{ 1 & (1 + e^{2\pi
i\over 3})& (1 + e^{-2\pi i\over 3})\cr -1 & 1& 1 \cr 1& (1 + e^{-2\pi
i\over 3}) & (1 + e^{2\pi i\over 3})\cr }, & \cr
\noalign {\hbox{
and }} & & (4.2.4c) \cr
 q_{\nu^\prime \nu} &= {1\over 3} \pmatrix{
1 & -1 & 1 \cr (1 + e^{-2\pi i\over 3}) & 1 & (1 + e^{2\pi i\over
3})\cr (1 + e^{2\pi i\over 3}) & 1 & (1 + e^{-2\pi i\over 3}) \cr }. &
\cr }$$ 

Using the symmetries studied in section 3, it is simple to show which
terms on the r.h.s. of eq.(4.2.1a) are equal. Besides, taking the
same steps as in section 4.1, we get that the terms [13]:
$<{\cal E}(0) {\cal E}(1) {\cal T}^-(2)>$, 
$<{\cal E}(0) {\cal T}^-(1){\cal T}^-(2)>$, 
$<{\cal E}(0) {\cal T}^-(1) {\cal U}(2)>$,  
$<{\cal E}(0){\cal U}(1) {\cal T}^-(2)>$, \break
$<{\cal T}^-(0) {\cal T}^-(1) {\cal T}^-(2)>$, 
$<{\cal T}^-(0) {\cal T}^-(1) {\cal T}^+(2)>$, 
$<{\cal T}^-(0){\cal T}^-(1) {\cal U}(2)>$ and \break
$<{\cal T}^-(0) {\cal U}(1) {\cal U}(2)>$ vanish identically. \par 

Taking into account the symmetries of section 3 and the fact that the
mentioned terms are zero, we get that ${\rm Tr}[{\bf K}^3]$ is equal
to \par 

$$\eqalignno{ {\rm Tr}[{\bf K}^3] &= <{\cal E}(0) {\cal E}(1) {\cal
E}(2)> + 3 <{\cal E}(0) {\cal E}(1) {\cal U}(2)> + 6 <{\cal E}(0)
{\cal T}^-(1) {\cal T}^+(2)> + & \cr
%
%segunda linha
%
 &+ 3<{\cal
E}(0) {\cal U}(1) {\cal U}(2)> + 6 <{\cal T}^-(0) {\cal T}^+(1) {\cal
U}(2)> + <{\cal U}(0) {\cal U}(1) {\cal U}(2)>. & \cr & & (4.2.5) \cr
}$$ 

All the terms on the r.h.s. of eq.(4.2.5) can be written as the sum of
multivariable Grassmann integrals. These integrals are reduced to five
different types of integrals. We will not detail the calculation of
these integrals,however, for it follows the steps presented in section
4.1 to derive the results of the integrals ${\cal H}_1(l), {\cal H}_2
(l,k)$ and ${\cal H}_3(l, k)$. \par 

The five types of integrals that appear in ${\rm
Tr}[{\bf K}^3]$ and their respective results are [14]: \par 

\noindent 1) \par 

\vskip -0.5cm

$$ {\cal G}_1 (l) \equiv \int \prod_{I=1}^{3N} \, d\eta_I
d\bar{\eta}_I \;\; e^{\sum\limits_{I,J= 1}^{3N} \bar{\eta}_I\; A_{I J}
\; \eta_J} \;\; \bar{\eta}_l \;\eta_l \;\; = 2^{N-1}, \eqno(4.2.6a) $$

\noindent for $ l= 1, 2, \cdots, N$. \par

\noindent 2) \par 

\vskip -0.5cm

$$\eqalignno{ {\cal G}_2 (l,k) &\equiv \int \prod_{I=1}^{3N} \,
d\eta_I d\bar{\eta}_I \;\; e^{\sum\limits_{I,J= 1}^{3N} \bar{\eta}_I\;
A_{I J} \; \eta_J} \;\; \bar{\eta}_l\; \eta_l \;\;\bar{\eta}_{N+k}
\;\eta_{N+k} & \cr
 & & \cr
%
%outras linhas
%
 & =\cases{ 2^{N-1}, &
$ k=l, \hskip 0.5cm l= 1, 2, \cdots, N$ \cr 2^{N-2}, & $ k\not=
l,\hskip 0.5cm l, k=1, 2, \cdots, N.$\cr } & (4.2.6b) \cr }$$

\noindent 3) \par

\vskip -0.5cm 

$$ \eqalignno{ {\cal J}_1 (l,k) &\equiv \int \prod_{I=1}^{3N} \,
d\eta_I d\bar{\eta}_I \;\; e^{\sum\limits_{I,J= 1}^{3N} \bar{\eta}_I\;
A_{I J} \; \eta_J} \;\; \bar{\eta}_l \;\eta_{l+1}
\;\;\bar{\eta}_{N+k}\; \eta_{N+k-1} & \cr
 & & \cr
%
% outras linhas
%
& \cases{ 2^{N-2}, & $ k= l+1, \hskip 0.5cm l= 1, 2, \cdots, N $ \cr
	     0, & $ k\not= l+1.$ \cr } & (4.2.6c) \cr }$$ 

\noindent 4) \par

\vskip -0.5cm 

$$ \eqalignno{ {\cal J}_2 (l_1, l_2, l_3) & \equiv \int
\prod_{I=1}^{3N} \, d\eta_I d\bar{\eta}_I \;\; e^{\sum\limits_{I,J=
1}^{3N} \bar{\eta}_I\; A_{I J} \; \eta_J} \;\; \bar{\eta}_{l_1}
\;\eta_{l_1} \;\;\bar{\eta}_{N+l_2} \; \eta_{N+l_2+1}\;\;
\bar{\eta}_{2N+l_3} \; \eta_{2N+l_3 -1} & \cr
 & & \cr
%
%outras linhas
%
 & = \cases{ 2^{N-3}, & $l_1 \not= l_2$, $l_1\not= l_3$ and
$l_3=l_2 +1$\cr 2^{N-2}, & $l_3=l_1$ and $l_2 = l_1-1$ \cr 0, & for
any other case. \cr } & (4.2.6d) \cr }$$

%\vfill
%\eject

\noindent 5) \par 

\vskip -0.5cm

$$\eqalignno{ {\cal G}_3 (l_1, l_2, l_3) &\equiv \int \prod_{I=1}^{3N}
\, d\eta_I d\bar{\eta}_I \;\; e^{\sum\limits_{I,J= 1}^{3N}
\bar{\eta}_I\; A_{I J} \; \eta_J} \;\; \bar{\eta}_{l_1} \;\eta_{l_1}
\;\;\bar{\eta}_{N+l_2}\; \eta_{N+l_2} \;\; \bar{\eta}_{2N+l_3} \;
\eta_{2N+l_3} & \cr
 & & \cr
%
% outras linhas
%
 & =\cases{2^{N-3}, &
$l_1\not= l_2 \not= l_3$ \cr 2^{N-2}, & $l_1= l_2 \not= l_3, l_1=l_3
\not= l_2, l_2=l_3 \not= l_1$ \cr 2^{N-1}, & $ l_1 = l_2 = l_3$. \cr }
& (4.2.6e) \cr }$$ 

\vskip 0.5cm

Using the results (4.2.6) to calculate the terms on the r.h.s. of
eq.(4.2.5), after some algebraic manipulation, we get \par 

$$\eqalignno{ {\rm Tr}[{\bf K}^3] &= 2^{2N} \big[ N^3 \Big(
 (E_0 - \mu)^3 + {3\over 4} (E_0 - \mu)^2 U +
 {3\over 16} (E_0 - \mu) U^2 + {1\over 64} U^3 \Big) + & \cr
%
%segunda linha
%
 & + N^2 \Big( 
{3\over 2} (E_0 - \mu)^3 + {15\over 8} (E_0 - \mu)^2 U + 
{3\over 2} (E_0 - \mu) \lambda_B^2 + 3 (E_0 - \mu) t^2 + & \cr
%
%terceira linha
%
 & \hskip 3cm +{15\over 16} (E_0 - \mu) U^2+ {3\over 4}t^2 U +
{3\over 8} U \lambda_B^2 + {9\over 64}U^3 \Big) + & \cr
%
%quarta linha
%
 &+ N
\Big( {3\over 8} (E_0 - \mu)^2 U +{3\over 8} (E_0 - \mu) U^2+ 
{3\over 2} (E_0 - \mu) t^2 -{3\over 8} U \lambda_B^2 + {3\over 4} t^2 U +
{3\over 32} U^3 \Big) \big]. & \cr & & (4.2.7) \cr }$$ 

\vskip 1cm 

\noindent {\bf 5. The Grand Canonical Partition Function for the
One--Dimensional Hubbard Model Up to Order $\beta^3$} 

\bigskip 

In eq.(2.4), we write the expansion of the grand canonical
partition function in the high temperature limit in terms of ${\rm
Tr}[{\bf K}^n]$, namely \par 

$$ \eqalignno{ {\cal Z}(\beta, \mu) &= {\rm Tr}( e^{- \beta {\bf K}})
& \cr
 &= {\rm Tr}[ 1\!\hskip -1pt{\rm I} - \beta {\bf K}] +
{\beta^2 \over 2} {\rm Tr}[ {\bf K}^2] - {\beta^3 \over 3!} {\rm Tr}[
{\bf K}^3]  + \cdots, & (5.1) \cr }$$ 

In reference [9], we calculated ${\rm Tr}[\lbarra - {\bf K}]$ for the
unidimensional periodic Hubbard model for a chain with $N$ space
sites, where $N$ can assume any arbitrary value, \par 

$$ {\rm Tr}[\lbarra - {\bf K}] = 2^{2N} \big[ 1 - N \beta \big( (E_0 -
\mu) + {U\over 4} \big) \big]. \eqno(5.2) $$ 

From expression (4.1.21), we have that \par 

$$\eqalignno{ {\rm Tr} [{\bf K}^2] &= 2^{2N} \big[ N^2 \big(
 (E_0 - \mu)^2 + {1 \over 2} (E_0 - \mu) U + {U^2 \over 16} \big) + & \cr
%
%segunda linha
%
 & \hskip 1.5cm + {N\over 2} \big( (E_0 - \mu)^2 + (E_0 - \mu) +
\lambda_B^2 + 2 t^2 + {3 \over 8} U^2 \big) \big]. & (5.3) }$$ 

In section 4.2, eq.(4.2.7), we got \par 

$$\eqalignno{ {\rm Tr}[{\bf K}^3] &= 2^{2N} \big[ N^3 \Big(
 (E_0 - \mu)^3 + {3\over 4} E_0 - \mu)^2 U +
 {3\over 16} (E_0 - \mu) U^2 + {1\over 64} U^3 \Big) + & \cr
%
%segunda linha
%
 & + N^2 \Big( {3\over 2} (E_0 - \mu)^3 + {15\over 8} (E_0 - \mu)^2 U + 
{3\over 2} (E_0 - \mu) \lambda_B^2 + 3 (E_0 - \mu) t^2 + & \cr
%
%terceira linha
%
& \hskip 3cm +{15\over 16} (E_0 - \mu) U^2+ {3\over 4}t^2 U + 
{3\over 8} U \lambda_B^2 + {9\over 64}U^3 \Big) + & \cr
%
%quarta linha
%
 &+ N
\Big( {3\over 8} (E_0 - \mu)^2 U +{3\over 8} (E_0 - \mu) U^2+ {3\over
2} (E_0 - \mu) t^2 -{3\over 8} U \lambda_B^2 + {3\over 4} t^2 U +
{3\over 32} U^3 \Big) \big]. & \cr & & (5.4) \cr }$$ 

Substituting (5.2--4) in (5.1), we get the grand canonical partition
function ${\cal Z}(\beta; \mu)$ of the unidimensional periodic Hubbard
model. However, we will use the Helmholtz free energy ${\cal
W}(\beta;\mu)$ to calculate physical quantities. The relation between
${\cal Z}(\beta; \mu)$ and ${\cal W}(\beta;\mu)$ is given by \par 

\vskip -0.5cm

$$ {\cal W}(\beta;\mu) = \ln {\cal Z}(\beta; \mu). \eqno(5.5) $$ 

The Helmholtz free energy of the one--dimensional periodic Hubbard
model, up to order $\beta^3$, is \par 

\vskip -0.5cm

$$\eqalignno{ {\cal W}(\beta;\mu) &= N \Big[ 2 \ln 2 + & \cr
%
%segunda linha
%
 & + \big( -{U^3\over 64} + {1\over 16} U \lambda_B^2
- {1\over 16} (E_0 - \mu)^2 U -{1\over 4} (E_0 - \mu) t^2 - {1\over
16} (E_0 - \mu) U^2 - {1\over 8} U t^2 \big) \beta^3 + & \cr
%
%terceira linha
%
 &  \hskip -0.5cm
+ \big( {1\over 4} (E_0 - \mu)^2 + {1\over 4}
\lambda_B^2 + {1\over 4} (E_0 - \mu) U + {t^2 \over 2} + {3\over 32}
U^2 \big) \beta^2 - \big( (E_0 - \mu) + {U\over 4} \big) \beta \Big] +
{\cal O}(\beta^4). & \cr & & (5.6) \cr }$$

\noindent The coefficients of the terms $\beta, \beta^2$ and
 $\beta^3$ in the expansion of ${\cal W}(\beta;\mu)$ are exact. \par 

Once we have the Helmholtz free energy in the high temperature limit,
we can derive any physical quantity in this limit. Here, we calculate
only three quantities: \par 

\vskip 0.5cm

\noindent $i)$ average energy per site: $<h>$. \par 

The simplest way to derive the average energy per site from the
Helmholtz free energy, is to scale the constants: $(E_0, t, U,
\lambda_B) \rightarrow (\alpha E_0, \alpha t, \alpha U,\alpha
\lambda_B)$ and substitute in eq.(2.1.2) to obtain 
${\cal W}(\beta; \mu; \alpha)$. \par 

$$ <h> (\beta) = - {1\over \beta N} {\partial {\cal W}(\beta;\mu; \alpha)
\over \partial \alpha} \Big|_{\alpha=1}. \eqno(5.7) $$ 

From eq.(5.6), we get that \par 

$$ \eqalignno{ <h> (\beta) & = E_0 + {U\over 4} + & \cr
%
%segunda linha
%
 & + \big[ - t^2 -{3\over 16} U^2 -{1\over 2} E_0 U + {1\over
4}U \mu - {1\over 2} E_0^2 + {1\over 2} E_0 \mu - {1\over 2}
\lambda_B^2 \big] \beta + & \cr
%
%terceira linha
%
 & \hskip -0.5cm +
\big[ {3\over 8} t^2 U - {1\over 2} t^2 \mu + {3\over 16} E_0 U^2 -
{1\over 8} U^2 \mu - {3\over 16} U \lambda_B^2 + {3\over 16} U E_0^2 -
{1\over 4} U E_0 \mu + & \cr
%
%quarta linha
%
 & \hskip 0.5cm +
{1\over 16} U \mu^2 + {3\over 64} U^3 + {3\over 4} t^2 E_0 \big]
\beta^2 + {\cal O}(\beta^3). & (5.8) \cr }$$ 

\vskip 0.5cm 

\noindent $ii)$ difference between average numbers of spin up and spin
down particles per site: $< n_\uparrow> - <n_\downarrow>$. \par 

From the definition of the Helmholtz free energy (eq.(5.5)), we have
that \par 

$$ < n_\uparrow> (\beta) - <n_\downarrow> (\beta) = -{1 \over \beta N}
{ \partial {\cal W}(\beta;\mu) \over \partial \lambda_B}. \eqno(5.9)
$$ 

Up to order $\beta^2$, we get from eq.(5.6) that, \par 

$$ < n_\uparrow> (\beta) - <n_\downarrow> (\beta) = {\lambda_B \over
8} ( U \beta + 4) \beta + {\cal O}(\beta^3). \eqno(5.10) $$ 

\vskip 0.5cm 

\noindent $iii)$ average of the square of the magnetization per site:
\par 

$$ \eqalignno{ < m_z^2> (\beta) &= {\lambda_B \over B^2} <({\bf n}_{i
\uparrow} - {\bf n}_{i \downarrow})^2 > & \cr
%
%segunda linha
%
 & =
\big( {1\over 2} g \mu_B \big)^2 {1\over \beta N} \Big[ {\partial
{\cal W}(\beta;\mu) \over \partial \mu} + 2 {\partial {\cal
W}(\beta;\mu) \over \partial U} \Big], & (5.11) \cr }$$ 

\noindent where $B$ is the external magnetic field. \par 

From eq.(5.6), we obtain that \par 

$$\eqalignno{ &< m_z^2> (\beta) = {1\over 4} g^2 \mu_B^2 \Big [
{1\over 2} + {1\over 8} U \beta + & \cr
%
%segunda linha
%
 & + \big[ -
{1\over 8} U E_0 + {1\over 8} \mu U - {1\over 32} U^2 + {1\over 8}
\lambda_B^2 - {1\over 8} E_0^2 + {1\over 4} E_0 \mu - {1\over 8} \mu^2
\big] \beta^2 \Big] + {\cal O}(\beta^3), & (5.12) \cr }$$ 

\noindent where {\it g} is the Land\'e's factor and $\mu_B$ is the
Bohr's magneton. \par 

\vskip 1cm 

\noindent {\bf 6. Conclusions} 

\bigskip 

Using the grassmannian nature of the fermionic fields, we extend the
results of our previous work [9] to higher order terms of
the expansion of the grand canonical partition function for 
unidimensional self--interacting fermionic models in the high
temperature limit. \par 

We are able to write the path integral for such models as a sum of
co--factors of a suitable matrix, the entries of which are commuting
quantities. This approach avoids ambiguities like the ones yielded
by Hubbard--Stratonovich transformation [3] for fermionic path 
integral. \par 

We apply the developed method to the Hubbard model in one--space
dimension with periodic boundary condition. We get the exact
coefficients of the $\beta-$expansion $( \beta = {1\over kT})$ of the
Helmholtz free energy up to order $\beta^3$ for a ring with $N$
space sites, where $N$ is arbritrary. \par 

The expansion (2.4) of the operator $ e^{-\beta {\bf K}}$ is valid for
any value of $\beta$. However, a sound physical meaning can only be
granted to its lower-order terms if it is assured that the expansion
converges. For finite values of the constants, there exists such a
range: the high temperature limit. \par 

The calculation of the coefficients of $\beta^4$ and $\beta^5$ terms
of the grand canonical partition function for the one--dimensional
Hubbard is to appear soon. \par 

The method presented here can be applied to any one--dimensional
self--interacting fermi\-onic model. We believe that this method can
be extended to higher space dimensions. \par

\vfill 
\eject

% Inicio do Apendice A 

\centerline{\bf Appendix A} \par 

\centerline{\bf Moments of Gaussian Grassmann Multivariable Integrals}
\par 

\bigskip 

\baselineskip=18pt 

It is a known result [11] for a Grassmann algebra of dimension
$2^{2N}$, composed of the generators $\{ \eta_1, \cdots, \eta_N;
\bar \eta_1, \cdots, \bar \eta_N\}$, that \par 

$$ \int \prod_{i = 1}^{\rm N} d\eta_i d\bar\eta_i \hskip 3pt
e^{\sum\limits_{ i, j = 1}^ {\rm N} \bar\eta_i A_{i j} \eta_j} = det
({\bf A}), \eqno(A.1) $$ 

\noindent where $A_{i j}$ are the entries of matrix {\bf A} and are
commuting quantities. \par 

We will show in this Appendix that the moments of integral (A.1) are
co--factors of {\bf A}. \par 

\vskip 0.5cm 

We first consider the case where we have one product $\bar\eta_l
\eta_k$ in the integrand of the gaussian integral (A.1), that is, \par

$$ M(l,k) \equiv \int \prod_{i = 1}^{\rm N} d\eta_i d\bar\eta_i
 \;\bar\eta_l \eta_k \hskip 3pt
 e^{ \sum\limits_{ i, j = 1}^{\rm N}\bar\eta_i A_{i j} \eta_j}, \eqno(A.2) $$ 

\noindent where $l, k$ are fixed and $ 1\leq l, k\leq N$. \par 

Due to the fact that for all Grassmann generators we have:
$\bar\eta_i^2 = \eta_i^2 = 0, i = 1, \cdots, N$, the only non--null
terms in eq.(A.2) are the ones where the integrand has \underbar{N}
products of the form: $\bar\eta_i \eta_j$. Eq. (A.2) becomes: \par 

$$ \eqalignno{
 M(l,k) = \int \prod_{i=1}^N \; d\eta_i d\bar\eta_i &
 \hskip 2pt \bar\eta_l \eta_k \hskip 3pt
{1 \over (N - 1)!}
\sum^N_{\textstyle {{ i_1, \cdots, i_{N-1} = 1} \atop {j_1, \cdots,
j_{N-1} = 1} } } A_{{i_1} {j_1}} \cdots A_{{i_{N-1}} {j_{N-1}}} \times
& \cr
%
%segunda linha da expressao
%
 & \times \bar\eta_{i_1}
\eta_{j_1} \; \bar\eta_{i_2} \eta_{j_2} \; \cdots \bar\eta_{i_{N-1}}
\eta_{j_{N-1}}, & (A.3) \cr } $$ 

\noindent and the indices are such that $ i_n \not= l, n=1, \cdots,
N-1$, and $ j_n \not= k, n=1, \cdots, N-1$. Once the product
$\bar\eta_{i_n} \eta_{j_n}$ is a commutative quantity, each term in
the sum of (A.3) appears $(N-1)!$ times. \par 

The $(N-1)!$ distinct terms in (A.3) can be generated by fixing one
configuration for $\{ i_1, i_2, \cdots, i_{N-1}\}$, for example, we
choose: $\{ i_1=1, \cdots, i_{l-1}=l-1, i_l= l+1, \cdots,
i_{N-1}=N\}$, and, taking all the terms coming from the sum over the
indices $j_n, n=1, \cdots, N-1$. Therefore, $M(l,k)$ becomes \par 

$$ \eqalignno{
 M(l,k) = \int \prod_{i=1}^N \; d\eta_i d\bar\eta_i &
 \hskip 2pt \bar\eta_l \eta_k \hskip 3pt
 \sum^N_{ j_1, \cdots,j_{N-1} = 1 \atop j_n \ne k }
 A_{1 {j_1}} \cdots A_{ l-1, j_{l-1}}
A_{ l+1, j_{l}} \cdots A_{N {j_{N-1}}} \times \cr
%
%segunda linha da expressao
%
 & \times \bar\eta_1 \eta_{j_1} \; \bar\eta_2 \eta_{j_2} \;
\cdots \bar\eta_{ l-1} \eta_{ j_{l-1}} \bar\eta_{ l+1} \eta_{j_l}
\cdots \bar\eta_N \eta_{j_{N-1}}. & (A.4) \cr } $$ 

Renaming the variables: $ j_l \rightarrow j_{l+1}, j_{l+1} \rightarrow
j_{l+2}, \cdots, j_{N-l} \rightarrow j_N$, we have that: \par 

$$ \eqalignno{
 M(l,k) = \int \prod_{i=1}^N \; d\eta_i d\bar\eta_i &
 \hskip 2pt \bar\eta_l \eta_k \hskip 3pt
\sum^N_{ j_1, \cdots,
j_{l-1} = 1 \atop j_{l+1}, \cdots, j_N = 1 }
 A_{1 {j_1}} \cdots A_{
l-1, j_{l-1}} A_{ l+1, j_{l+1}} \cdots A_{N {j_N}} \times \cr
%
%segunda linha da expressao
%
 & \times \bar\eta_1 \eta_{j_1} \;
\bar\eta_2 \eta_{j_2} \; \cdots \bar\eta_{ l-1} \eta_{ j_{l-1}}
\bar\eta_{ l+1} \eta_{j_{l+1}} \cdots \bar\eta_N \eta_{j_N}. & (A.5)
\cr } $$ 

Defining the matrix ${\bf B}(l,k)$ as: \par 

$$ B_{i j} (l,k) = \cases { A_{i j}, & if $ i\neq l$ and $j\neq k$ \cr
\delta_{il} \delta_{jk}, & if $i=l$ or $j=k$ \cr }, \eqno(A.6) $$ 

\noindent and $i, j= 1, 2, \cdots, N$. \par 

Using the definition of matrix ${\bf B}(l,k)$,the expression of
$M(l,k)$ is re--written as: \par 

$$ \eqalignno{
 M(l,k) = \int \prod_{i=1}^N \; d\eta_i d\bar\eta_i &
\hskip 3pt
 \sum^N_{ j_1, \cdots, j_N= 1 }
B_{1 {j_1}} \cdots B_{
l-1, j_{l-1}} B_{ l, j_l} \cdots B_{N {j_N}} \times \cr
%
%segunda linha da expressao
%
 & \times \bar\eta_1 \eta_{j_1} \; \cdots
\bar\eta_l \eta_{j_l} \; \cdots \bar\eta_N \eta_{j_N}. & (A.7) \cr }
$$ 

Integrating over $\bar\eta_i$, and using the definition of
determinant [10, 15], we finally have that \par 

$$ \eqalignno{ M(l,k) &= det {\bf B} & \cr &= (-1)^{l+k} A(l,k), &
(A.8) \cr } $$ 

\noindent where $A(l,k)$ is the minor determinant of matrix {\bf A},
when the line $l$ and the column $k$ are deleted. $M(l,k)$ is the
cofactor of matrix {\bf A}. \par 

\vskip 0.5cm 

Using an analogous procedure, we now consider the case of moments of
the gaussian Grassmann multivariable integral when we have $m$
products: \par 

$$ \bar\eta_{l_1} \eta_{k_1} \hskip 2pt \bar\eta_{l_2} \eta_{k_2}
\hskip 2pt \cdots \bar\eta_{l_m} \eta_{k_m} $$ 

\noindent in the integrand of (A.2), where $ m\leq N$. \par 

Consider the fixed sets: $ L= \{ l_1, l_2, \cdots, l_m\}$ and $K=\{
k_1, k_2, \cdots, k_m \}$. We define $M(L,K)$ as \par 

$$ M(L,K) \equiv \int \prod_{i = 1}^{\rm N} d\eta_i d\bar\eta_i
\;\bar\eta_{l_1} \eta_{k_1} \cdots \bar\eta_{l_m} \eta_{k_m} \
\hskip 3pt
 e^{ \sum\limits_{ i, j = 1}^{\rm N} \bar\eta_i A_{i j} \eta_j},
\eqno(A.9) $$ 

\noindent and the products are ordered such that: $ l_1< l_2< \cdots <
l_m$ and $k_1< k_2< \cdots< k_m$. \par 

Using an analogous reasoning, we obtain \par 

$$ \eqalignno{
 M(L,K) &= \int \prod_{i=1}^N \; d\eta_i d\bar\eta_i
\hskip 3pt
 \sum^N_{ j_1, \cdots, j_N= 1 }
  B_{1 {j_1}} \cdots B_{
l-1, j_{l-1}} B_{ l, j_l} \cdots B_{N {j_N}} \times \cr
%
%segunda linha da expressao
%
 & \times \bar\eta_1 \eta_{j_1} \; \cdots
\bar\eta_l \eta_{j_l} \; \cdots \bar\eta_N \eta_{j_N} = det{\bf
B}(L,K), & \cr
 &=(-1)^{(l_1 + l_2+ \cdots+ \l_m) + (k_1+ k_2+
\cdots+ k_m)} A(L,K), & (A.10)\cr } $$ 

\noindent where the matrix ${\bf B}(L,K)$ is defined as: \par 

$$ B_{i j} (L,K) = \cases { A_{i j}, & if $ i\neq l_1, \cdots, l_n \;$
and $ \; j\neq k_1, \cdots k_n$ \cr & \cr \delta_{i {l_1}} \delta_{j
{k_1}}, & if $i=l_1 $ or $j=k_1$ \cr \vdots & \cr \delta_{i {l_m}}
\delta_{j {k_m}}, & if $i=l_m $ or $j=k_m$, \cr } \eqno(A.11) $$ 

\noindent and $ i,j =1, 2, \cdots, N$. $A(K,L)$ is the determinant of
the matrix obtained from matrix {\bf A} by deleting the lines: $\{
l_1, l_2, \cdots, l_n\}$, and, the columns: $\{ k_1, k_2, \cdots,
k_n\}$. \par 

\vskip 0.5cm 

In summary, we can say that the effect of the presence of a product
$\bar\eta_l \eta_k$ within the integrand of the gaussian integral
(A.2), is to replace the line $l$ of matrix {\bf A}, $A_{l j}$, by
$\delta_{j k}$, and its column $k$, $A_{i k}$, by $\delta_{i l}$. In
their turn, the determinants of matrices ${\bf B}(L,K)$, eq.(A.10),
are easily written in terms of determinants of matrices of smaller
dimension. Hence products of Grassmann generators cut down the
dimension of the matrices the determinant of which we are to
calculate.

% Termino do Apendice A 

\bigskip 

\noindent {\bf Acknowledgements} \par 

\bigskip 

I.C.C, E.V.C.S. and S.M. de S. thank CNPq for financial support.
M.T.T. thanks CNPQ and FINEP for partial financial support.

%\vfill
%\eject 

\vskip 1cm 

% Inicio das Refrencias 

\centerline {\bf REFERENCES} \par
\bigskip 

\item{1.}{ J. Hubbard, Proc. Roy. Soc. {\bf A277} (1963) 237; {\bf
A281} (1964) 401; } \par 

\item{2.}{ J. Hubbard, Phys. Rev. Lett. {\bf 3} (1959) 77; \hfill
\break R.L. Stratonovich, Sov. Phys. Dokl. {\bf 2} (1958) 416;
\hfill\break J.W. Negele and H. Orland, {\it Quantum Many--Particle
Systems}, Addison-Wesley, Reading, MA (1988);} \par 

\item{ 3.}{ W.E. Evenson, J.R. Schrieffer and S.Q. Wang, J. Appl.
Phys. {\bf 41} (1970) 1199;  \hfill \break
 H. Hasegawa, Journ. of the Phys. of Soc. of Jap. {\bf 46} (1979)
1504;  \hfill \break
 J. Hubbard, Phys. Rev. {\bf B19} (1979) 2626;}\par 

\item{ 4.}{ W.O. Putikka, M.U. Luchini and T.M. Rice, Phys. Rev. Lett.
{\bf 68} (1992) 538; L. Chen, C. Bourbannais, T. Li and A--M S.
Tremblay, Phys. Rev. Lett. {\bf 66} (1991) 369; C.J. Thompson, Y.S.
Yang, A.J. Guttmann and M.F. Sykes, J. Phys. {\bf A24} (1991) 1261;
K. Kubo and M. Tada, Prog. Theoret. Phys. {\bf 69} (1983) 1345, {\bf 71}
(1984) 479 ; W. Brauneck, Z. Physik {\bf B28} (1977) 291;} \par 

\item{ 5.}{ E.H. Lieb and F.Y. Wu, Phys. Rev. Letters {\bf 20} (1968)
1445; \hfill \break
Y. Nagaoka, Solid State Commun. {\bf 3} (1965) 409; Phys. Rev. 
{\bf 147} (1965) 392; \hfill \break
M. Takahashi, Prog. Theoret. Phys. (Kyoto) {\bf 43} (1970) 1619;} \par 

\item{ 6.} { M. Takahashi, Prog. Theor. Phys. {\bf 47} (1972) 69;}
\par 

\item{ 7.}{ H. Shiba and P.A. Pincus, Phys. Rev. {\bf B5} (1972)
1966;} \par 

\item{ 8.}{ S.M. de Souza and M.T. Thomaz, J. Math. Phys. {\bf 32}
(1991) 3455;} 

\item{ 9.}{ I.C. Charret, E.V. Corr\^ea Silva, S.M. de Souza and M.T.
Thomaz, J. Math. Phys. {\bf 36} (1995) 4100}; 

\item{ 10.}{ S.M. de Souza and M.T. Thomaz, J. Math. Phys. {\bf 31}
(1990) 1297;} 

\item{ 11.} { C. Itzykson and J.--B. Zuber, {\it Quantum Field
Theory}, McGraw--Hill (1980); \hfill\break U. Wolf, Nucl. Phys.
{\bf B225} [FS9]  (1983) 391; } \par 

\item{ 12.}{ We are using the convention: $\sigma= 1 = \uparrow$ 
and $\sigma= -1 = \downarrow$; }  \par

\item{ 13.} {  We continue to use the notation (4.1.5);  }\par

\item { 14.} { Due to the fact that {\bf A} is a
block--matrix, the contributions from the sectors $\sigma\sigma =
\uparrow\uparrow$ and $\sigma\sigma = \downarrow\downarrow$ are
decoupled. Because ${\bf A}^{\uparrow\uparrow} =
 {\bf A}^{\downarrow\downarrow}$, the results of the integrals in the two
sectors $\sigma\sigma = \uparrow\uparrow$ and $\sigma\sigma =
\downarrow\downarrow$ are equal. That is why it is enough to calculate
the integrals in only one sector; }\par

\item{ 15.} { D. Kreider et al, {\it An Introduction to Linear
Analysis}, Addison--Wesley, Ontario (1966).} \par 

% Termino das Referencias 

\vfill \eject 

\bye